\documentclass{scrartcl}

\usepackage[T1]{fontenc}
\usepackage[utf8]{inputenc}
\usepackage{graphicx}
\usepackage{mathtools}
\usepackage{amssymb}
\usepackage[sort,longnamesfirst]{natbib}
\usepackage{hyperref}

\DeclarePairedDelimiter{\braces}{\lbrace}{\rbrace}
\DeclarePairedDelimiter{\parens}{\lparen}{\rparen}

\title{MCMC Methods for Multi-modal Distributions}
\subtitle{Appears in ``Handbook of Markov Chain Monte Carlo", 2nd edition.}
\author{
  Krzysztof Łatuszyński \\
  \and
  Matthew T. Moores \\
  \and
  Timoth{\'e}e Stumpf-F{\'e}tizon
}

\begin{document}

\maketitle

Three and a half billion years of evolution and no animal has wheels?! "The wheel may be one of those cases where the engineering solution can be seen in plain view, yet be unattainable in evolution because it lies the other side of a deep valley, cutting unbridgeably across the massif of Mount Improbable" explained the evolutionary biologist Richard Dawkins in a popular 1996 article \citep{dawkins_wheels}. Indeed, multimodal distributions are hard for local samplers. So, how are the engineers of today doing, and are we at the 4000 BCE mark yet, when the wheel was invented in Mesopotamia? 

In this chapter, we explain the fundamental challenges of sampling from multimodal distributions, particularly for high-dimensional problems. We present the major types of MCMC algorithms that are designed for this purpose, including parallel tempering, mode jumping and Wang--Landau, as well as several state-of-the-art approaches that have recently been proposed. We demonstrate these methods using both synthetic and real-world examples of multimodal distributions with discrete or continuous state spaces.

\section{Introduction}
\label{sec:intro}

Multimodal posterior distributions are common in applications. They arise naturally in many problems across sciences; in biology \citep{ballnus2017comprehensive}, genetics \citep{bouckaert2019beast}, astrophysics \citep{feroz2009multinest,feroz2013importance}, statistical physics \citep{kamberaj2020molecular} or signal processing \citep{ihler2004nonparametric}. They are also present in certain statistical models due to their inherent problems with identifiability, such as label switching in Bayesian mixture models \citep{MR2182987}, or parameter constraints in the curved exponential family of models \citep{Sundberg2010}. Multimodality can also result from model misspecification or parameter identifiability issues even in situations where it is not inherent to the inference problem at hand \citep{Drton2004}. Generally speaking, any problem that leads to multiple local maxima in frequentist maximum likelihood inference, or multiple local minima in empirical error minimization, will result in a multimodal posterior distribution for its Bayesian counterpart, at least for some choice of the prior distributions. Multimodal distributions arise also in various areas of machine learning, such as fitting and uncertainty quantification for neural networks \citep{MR4444376, MR4675034}.

Exploring multimodal distributions is very challenging and poor mixing of standard Markov chain Monte Carlo (MCMC) methods on such targets is notorious. Due to their dynamics, these algorithms struggle with crossing low-probability barriers separating the modes and, consequently, take a long time before they find the modes or move from one mode to another, even in low dimensions. In this chapter, we discuss the difficulties of multimodal sampling in more detail and present a number of classical and recent MCMC algorithms designed specifically for this problem. We illustrate their applicability and performance on a range of synthetic and real examples.

\section{Notation and Preliminaries}
\label{sec:prelim}

We start by fixing notation and introducing a few Markov chain concepts that will help us understand the properties of multimodal distributions and the Markov chains that target them. For more complete background on Markov chains theory, we refer to \citep{MR2095565, MR1287609, MR3889011} as well as other chapters in this volume.

Let $\pi$ be the multimodal target distribution defined on $(\mathcal{X}, \mathcal{B}(\mathcal{X}))$. The state space $\mathcal{X}$ will often be high dimensional, however its exact nature can vary. It can be continuous, such as $\mathcal{X} \subseteq \mathbb{R}^{d}$, discrete, such as $q$ colors on $M$ sites, $\mathcal{X} = \braces{1,2, \dots, q}^{M}$, or mixed, such as in the Kingman coalescent model \citep{MR0671034}, where both the tree topology and branch lengths are encoded. We shall often abuse notation by writing $\pi$ for both, the target distribution and its density with respect to a suitable reference measure - the meaning shall be clear from the context. In Bayesian statistics, $\pi$ is the posterior distribution on a parameter space $\mathcal{X} = \Theta$. From a prior $p(\theta)$ and a likelihood function $p(\mathrm{data} | \theta)$, Bayes' theorem yields the posterior distribution    
\begin{equation*}
    \pi(\theta) = p(\theta | \mathrm{data}) = \frac{p(\mathrm{data} | \theta) p(\theta)}{\mathcal{Z}(\mathrm{data})},
\end{equation*}
where $\mathcal{Z}(\mathrm{data}) = \int_{\Theta} p(\mathrm{data} | \theta) p(\theta) d\theta$ is the normalizing constant. We presume that multimodality in the posterior is induced by the likelihood, rather than the prior.

Inference problems under $\pi$ may be formulated as integrals of form
\begin{equation*}
    \pi(f) = \int_{\mathcal{X}} f(x) \pi(dx)
\end{equation*}
for a given test function $f$. We iteratively simulate an ergodic Markov chain $\braces{X_{n}}_{n=0}^{m}$ with invariant distribution $\pi$ and estimate $\pi(f)$ by the \emph{ergodic average}
\begin{equation} 
    \label{eq:erg_av}
    \widehat{\pi(f)}_{m} = \frac{1}{m+1} \sum_{n=0}^{m} f(X_{n}).
\end{equation}
Such a Markov chain is constructed from a Markov transition kernel $P$ that, given the initial distribution $X_{0} \sim \mu_{0}$, defines the dynamics of $\braces{X_{n}}_{n=0}^{\infty}$ by letting $X_{n+1}|X_{n} = x \sim P(x, \cdot)$, where for every $x \in \mathcal{X}$ the object $P(x, \cdot)$ defines a probability distribution. It is convenient to write the distribution of $X_{n}$ in terms of iterates of the Markov transition kernel $P$ by noticing that it acts on probability measures as follows. For a probability measure $\mu_{0}$, define
\begin{equation}
    \label{eq:nuP}
    \mu_{0} P(A) = \int_{\mathcal{X}} P(x, A) \mu_{0}(dx) = \mathbb{P}(X_{1} \in A|X_{0} \sim \mu_{0}).
\end{equation}
Hence $\mu_{0} P = \mathcal{L}(X_{1}|X_{0} \sim \mu_{0})$, and by iterating \eqref{eq:nuP},   
\begin{equation}
    \label{eq:kernel_iterate}
    \mu_{0} P^n = (\mu_{0} P^{n-1})P = \mathcal{L}(X_{n} | X_{0} \sim \mu_{0}).
    \end{equation}
To ensure asymptotic validity of this procedure, $P$ must have $\pi$ as its invariant distribution, that is satisfy $\pi P = \pi$, and for any starting distribution $\mu_{0}$ the kernel iterate $\mu_{0} P^n$ must converge to $\pi$ in a suitable norm, such as the 
\emph{total variation norm}. For two probability measures $\nu$ and $\mu$, the total variation norm is defined as 
\begin{equation}
    \|\nu - \mu\|_{TV} = \sup_{A \in \mathcal{B}(\mathcal{X})} |\nu(A) - \mu(A)|.
\end{equation}
Furthermore, we say that the transition kernel $P$ is reversible with respect to $\pi$ if for every $A, B \in \mathcal{B}(\mathcal{X})$ it satisfies 
\begin{equation}
    \label{eq:rev}
    \int_{A} P(x, B) \pi(dx) = \int_{B} P(y, A) \pi(dy),
\end{equation}
for which the shorthand $\pi(x)P(x,y) = \pi(y)P(y,x)$ is often used. Reversibility implies that $\pi$ is invariant for $P$ and together with mild additional assumptions also yields convergence of $\mu_{0}P^{m} \to \pi$. Hence, the common strategy is to design $P$ such that reversibility holds, and the Metropolis-Hastings (MH) algorithm is a prime example of how to do that: given the current state $X_{n} = x$, an irreducible transition kernel $Q$ is used to generate a new point $y$ from the distribution $Q(x, \cdot)$. This proposal is then accepted, in which case $X_{n+1} \coloneq y$, or rejected, in which case $X_{n+1} \coloneq X_{n}$. The acceptance probability is 
\begin{align}
    \label{eq:acc1}
    \alpha(x,y) & = \min\braces*{1, \frac{\pi(y)q(y,x)}{\pi(x)q(x,y)}} \\ 
    \label{eq:acc2} 
    & = \min\braces*{1, \frac{\pi(y)}{\pi(x)}}, 
    \qquad \text{if} \quad q(y,x) = q(x,y),
\end{align}
where $y \rightarrow q(x,y)$ is the density of $Q(x, \cdot)$. Note that the function in \eqref{eq:acc1} is chosen precisely so that \eqref{eq:rev} is satisfied. While other acceptance functions are possible, and may become relevant in intractable likelihood scenarios \citep{gonccalves2023exact, MR4430963}, using them would not affect the issue of multimodality, so we restrict our attention to the Metropolis-Hastings acceptance function.  

While reversibility \eqref{eq:rev} yields a mechanism for constructing $P$ and gives insight into the one-step-dynamics of the Markov chain in question, the Kac Theorem \cite[Theorem 10.0.1]{MR1287609} gives insight into the long term dynamics through relative occupation measures of sets, without assuming reversibility. Let $A, B \in \mathcal{B}(\mathcal{X})$ and $\pi(A) > 0$, then
\begin{equation}
    \label{eq:Kac}
    \pi(B) = \int_{A} \mathbb{E}_{x} \sum_{i=1}^{\tau_{A}} \mathbb{I}_{B}(X_{i}) \pi(dx),
\end{equation}
where $\tau_{A} \coloneq \min\braces{i \geq 1: X_{i} \in A}$ is the first hitting time for set $A$ and $\mathbb{E}_x$ denotes expectation with respect to the distribution of the Markov chain trajectory given $X_{0} = x$.

For reliable MCMC estimation, as in \eqref{eq:erg_av}, it is desirable that the rate of convergence of $\mu_{0}P^{m}$ to $\pi$ is geometric \citep{MR1624414}. \emph{Geometric ergodicity} of a reversible Markov chain is related to $S(P)$, the spectrum of its transition kernel $P$ considered as an operator on $L_{0}^{2}(\pi)$ - the space of zero-mean, square integrable functions with respect to $\pi$. We refer to \citet{MR1888449} and \citet[Chapter 22]{MR3889011} for a more complete background. Briefly, the spectrum of $P$ is defined as
\begin{equation}
    \label{def:spectrum}
    S(P) \coloneq \braces{\lambda \in \mathbb{C}: P - \lambda I \; \text{is not invertible}},
\end{equation} 
and $P$ being a Markov transition kernel implies that $S(P) \subseteq \braces{\lambda \in \mathbb{C}: |\lambda| \leq 1}$. Additionally, if $P$ is reversible then $S(P)$ is real, i.e. $S(P) \subseteq [-1,1]$. The object of interest is the absolute spectral gap of $P$ defined as
\begin{equation} 
    \label{def:absgap}
    \operatorname{gap}(P) \coloneq 1 - \sup \braces{|\lambda|: \lambda \in S(P)}.
\end{equation} 
Note that if we considered $P$ on $L^{2}(\pi)$ - the space of square integrable functions with respect to $\pi$ that are not necessarily required to be zero-mean - then the constant function $g = 1$ would be the eigenfunction associated with the eigenvalue 1. However, nontrivial constant functions are not in  $L_{0}^{2}(\pi)$. Consequently, if as defined above, $\text{gap}(P) > 0$, then $P$ is geometrically ergodic \citep{MR1448322, MR2981426}, i.e. convergence in total variation satisfies
\begin{equation} 
    \label{eq:ge}
    \|P^{m}(x, \cdot) - \pi(\cdot)\|_{\text{TV}} \leq C_{\rho}(x) \rho^{m},
\end{equation}
for any $\rho \in (1-\text{gap}(P), 1)$ and some function $C_{\rho}: \mathcal{X} \to \mathbb{R}_{+}$ \cite[Section~6]{MR2114987}. Therefore, studying $\operatorname{gap}(P)$, and in particular how it scales with various parameters of the problem, gives insight into the performance of algorithms and their scaling. Such parameters include the dimension of the parameter space in the Bayesian setting (which becomes dimension of the state space of the MCMC used for inference), or the size of the data set used for inference.

Other tools for assessing and optimizing the performance of accept-reject MCMC algorithms rely on diffusion limits \citep{MR1428751, MR1888450}. In this technique, one considers what happens to the Markov chain as the problem dimension $d$ increases towards infinity, and the proposal variance, say $\sigma$, is scaled appropriately with $d$, so that the acceptance rate does not go to $0$, which would result in no mixing, nor to $1$, which would result in suboptimal mixing. Under suitable regularity conditions and time-space rescaling, the resulting limiting process is a continuous diffusion with an explicit speed function. Both - the scaling of $\sigma$ and the speed function - give insight into relative performance of algorithms and can be used for optimization of the algorithm's tuning parameters. In the instance of the \emph{Random Walk Metropolis} algorithm (RWM), the conclusions from the diffusion limit are that as dimension $d$ grows, the proposal must be scaled as $l/\sqrt{d}$, and the constant $l$ is such that the optimal average acceptance rate of $0.234$ is achieved. Analogous results also hold for other standard algorithms, such as Metropolis-adjusted Langevin algorithm (MALA), Hamiltonian Monte Carlo (HMC), pseudomarginal MCMC algorithms, and accept-reject algorithms that employ other acceptance functions. We refer to \cite{rosenthal2011optimal} in this volume for more background on diffusion limits and scaling.

\section{Fundamental Challenges of Multimodal Sampling}
\label{sec:challenges}

There are three fundamental challenges when sampling from a multimodal target distribution, and all three exacerbate with dimension. We elaborate on these challenges so that algorithms can then be evaluated in how well they address them.

\subsection{Moving Between the Modes}
\label{subsec:between_modes}

Many Markov chains are constructed by designing local moves, by which we mean that the distribution of $X_{n+1}|X_{n} = x$ has probability mass around $x$, and moves where $|X_{n-1} - X_{n}|$ is large, are unlikely. A RWM algorithm with proposal $Y_{n+1}|X_{n} = x \sim N(x, \Sigma)$ is one such example. Typically, the moves become more local as dimension increases. This can be seen from the diffusion limit argument in Section \ref{sec:prelim}, where the scale of the RWM proposal is of order $d^{-1/2}$ so as to keep the acceptance probability constant as $d \to \infty$. Analogous results with different rates exist for other algorithms as well \citep{rosenthal2011optimal}.

\begin{figure}
    \begin{center} 
        \includegraphics[scale=.66]{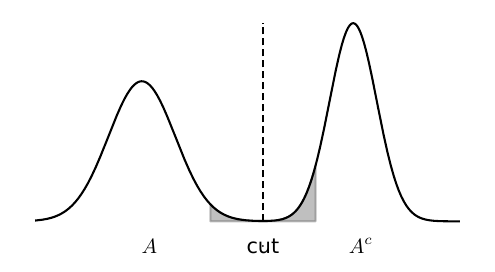}
        \includegraphics[scale=.66]{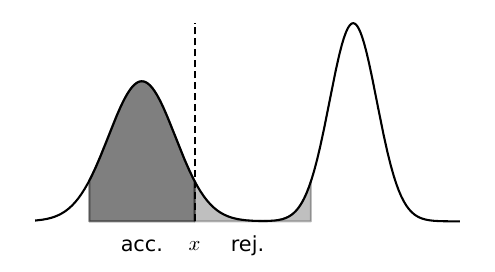}
    \end{center}
    \caption{Right - a proposal from location $x$ to its neighborhood on the left, where $\pi(y) > \pi(x)$, will be accepted with probability 1, but a proposal to its neighborhood on the right,  where $\pi(y) < \pi(x)$, will be rejected with positive probability. Moving between the modes via multiple local moves is very unlikely. Left - illustrates \eqref{eq:conductance_local} - the first integral, $\pi_{A}$-probability of being close to the cut, is the shaded area of $A$ and it becomes exponentially small in dimension of the state space or distance between the modes. So does the second integral, i.e. the probability of jumping from the unshaded area of $A$ to the set $A^{c}$.}
    \label{fig:acc-rej}
\end{figure}

Any algorithm that makes only local moves will struggle to cross low probability barriers and move between well separated modes. Going from one mode, where relative values of $\pi(x)$ are high, towards another mode across a valley, where $\pi(x)$ is small, would require a large number of consecutive downwards steps to occur. Taking the RWM as an example, the proposals $y$ are symmetric around the current state $x$ and accepted with probability \eqref{eq:acc2}. Therefore, if attempting to move down the valley where $\pi(y) < \pi(x)$, the probability of rejection is strictly positive. However, a proposal back towards the mode that we are hoping to leave is accepted with probability~1 since $\pi(y) > \pi(x)$. See the right panel of Figure \ref{fig:acc-rej} for an illustration. Hence, the consecutive accept-reject coins are biased towards going back to the current mode, and as the valley deepens - as it would with increasing dimension, or as the modes become more separated - traversing it becomes exponentially more difficult. Similar intuition can be derived for any local reversible Markov chain based on the detailed balance condition \eqref{eq:rev}. Moves towards neighboring sets where $\pi$ is smaller must be on average less likely than their reversals to compensate for the decreasing probability mass of the stationary distribution in~\eqref{eq:rev}.

These intuitive arguments help understand the challenge that multimodality is posing, but also can be made rigorous. To this end, for a $\pi$-invariant Markov transition kernel $P$ on $(\mathcal{X}, \mathcal{B}(\mathcal{X}))$, define conductance as
\begin{equation}
    \label{eq:conductance}
    \kappa \coloneq \inf_{A \in \mathcal{B}(\mathcal{X}): 0 < \pi(A) \leq 1}  \kappa(A) \stackrel{(*)}{=} \inf_{A \in \mathcal{B}(\mathcal{X}): 0 < \pi(A) \leq 1/2} \kappa(A),
\end{equation}
where 
\begin{equation}
    \label{eq:kA}
    \kappa(A) \coloneq \frac{\int_{A} P(x, A^{c}) \pi(dx)}{\pi(A) \pi(A^{c})} \stackrel{(*)}{=} \kappa(A^{c}),
\end{equation}
and where equalities marked with $(*)$ in \eqref{eq:conductance} and \eqref{eq:kA} hold for reversible chains only. Then, for a reversible Markov transition kernel $P$,
\begin{equation}
    \label{eq:cheeger_right}
    \frac{\kappa^{2}}{8} < 1 - \sup \braces{\lambda : \lambda \in S(P)} < \kappa.
\end{equation}
Consequently, recalling the definition of the absolute spectral gap \eqref{def:absgap}, it implies that
\begin{equation}
    \label{eq:cheeger_gap}
    \operatorname{gap}(P) < 1 - \sup \braces{\lambda : \lambda \in S(P)} < \kappa.
\end{equation}
We refer to \citet{MR0930082, jerrum1988conductance, chatterjee2023spectral} and \citet[Chapter 22]{MR3889011} for details on this result in various settings, including a version for nonreversible chains, and to \citet{mihail1989conductance, MR3726904} for a connection to mixing times. Combining \eqref{eq:cheeger_gap} with \eqref{eq:ge} yields a powerful tool. For example, showing that conductance decreases exponentially in dimension, or more generally, in problem size, implies that convergence time to stationarity is exponentially large in problem size. Such Markov chains are termed torpidly mixing, as opposed to rapidly mixing when this dependence is polynomial, and are considered not to scale with problem size (c.f. \cite{woodard2009sufficient}). To show torpid mixing of local Markov chains through conductance, one typically cuts the state space through the deepest point in the valley between the modes into $A$ and $A^{c}$, assuming without loss of generality that $\pi(A) \leq 1/2$. Then, one estimates conductance by
\begin{equation}
\begin{aligned}
    \kappa(A) 
    & \coloneq \frac{\int_{A}P(x, A^{c})\pi(dx)}{\pi(A)\pi(A^{c})} \leq \frac{2}{\pi(A)} \int_{A} P(x, A^{c}) \pi(dx) \\ 
    & = 2 \int_{A} P(x, A^{c}) \frac{\pi(dx)}{\pi(A)} = \spadesuit, 
\end{aligned}
\end{equation}
where $\pi(dx) / \pi(A)$ is the stationary distribution truncated to $A$, and
\begin{equation}
\begin{aligned}
    \label{eq:conductance_local}
    \spadesuit 
    & = 2\int_{A \cap \braces{\text{close to the cut}}} P(x, A^{c})\frac{\pi(dx)}{\pi(A)} + 2 \int_{A \cap \braces{\text{far from the cut}}} P(x, A^{c}) \frac{\pi(dx)}{\pi(A)} \\
    & < 2 \int_{A \cap \braces{\text{close to the cut}}} \frac{\pi(dx)}{\pi(A)} + 2 \int_{A \cap \braces{\text{far from the cut}}} P(x, A^{c}) \frac{\pi(dx)}{\pi(A)}. 
\end{aligned}
\end{equation}

Now, for many scenarios, for a suitable choice of the set $\{\text{close to the cut}\}$, it can be verified that the first integral is exponentially small because the probability mass close to the cut is exponentially small in $d$, and the second integral is exponentially small because $P(x, A^{c})$ is exponentially small in $d$ when $x$ is far from the cut. See the left panel of Figure~\ref{fig:acc-rej} for illustration. When scaling the distance between modes rather than the dimension of the state space, a similar analysis shows that conductance decreases exponentially in the distance between the modes. We refer to \citet[Proposition 2.1]{Rep_Att_HMC}, which can be adapted to obtain such a result for Hamiltonian Monte Carlo.  

Since the above conductance argument relies on reversibility, it is tempting to hope that nonreversible local algorithms perform better. However, it is the local proposal that is at the heart of the problem, not reversibility. To see this, once again cut the state space through the deepest point in the valley and let the two halves be $B^{*}$ and $C^{*}$ (rather than $A$ and $A^{c}$). Now, let $A^{*}$ be the boundary between $B^{*}$ and $C^{*}$. Clearly, $A^{*}$ is a lower dimensional manifold and $\pi(A^{*}) = 0$. However, in many multimodal problems, we can now ``inflate" $A^{*}$ so that its probability mass is exponentially small in dimension, but $A^{*}$ is ``thick" enough that crossing it in one nonreversible step is also exponentially unlikely. Let $A$ be this inflated $A^{*}$ and let $B = B^{*} \setminus A$ and $C = C^{*} \setminus A$. To conclude, recall the Kac Theorem \eqref{eq:Kac} with $A$ and $B$ constructed above. Since $\pi(A)$ is exponentially small and $\pi(B)$ is $O(1)$, the return time $\tau_{A}$ must be exponentially large to balance this and hence it takes exponential time to revisit $A$ and, equivalently, to move between $B$ and $C$.

Thus, moving between the modes is typically exponentially difficult for local algorithms, and in many scenarios verifiably exponentially difficult. Nonlocal moves that attempt to jump between the modes in one step seem to be a natural remedy. To construct such moves, a list of mode locations is needed, which we elaborate on in the next section. Even when such a list is available, jumping between the modes is still challenging. The generic way to ensure ergodicity is for the Markov transition kernel of the jump move to be designed as a mixture proposal and to satisfy the detailed balance condition \eqref{eq:rev} through the accept-reject mechanism \eqref{eq:acc1}. This puts the jump kernels into the independence sampler framework, where the spectral gap is characterized as follows \citep{MR2256484, MR2303970, liu1996metropolized}:
\begin{equation}
  \label{eq:jump_gap}
  \text{gap}(P) = 1/w^{*}, \qquad w^{*} = \operatorname*{ess\ sup}_{x \in \mathcal{X}} \braces*{\frac{\pi(x)}{Q_{J}(x)}}. 
\end{equation}
Here $Q_{J}(x)$ is the proposal density of the jump kernel, and the essential supremum is taken with respect to $\pi$. It is now clear that for the spectral gap to exist, the proposal kernel must have tails at least as heavy as the target. However, an exact match will usually be impossible and overdispersion will typically result in $w^{*}$ being exponentially small in dimension. To see this, simply set $\pi = N(0, I_{d})$ and $Q_{J}(x) = N(0, (1+\varepsilon) I_{d})$. Therefore, even if the location of the modes is known, a naive jump kernel might be useful when modes are well separated or the state space is of moderate dimension, but without further ideas it will quickly break down as dimension increases.

\subsection{Finding the Modes}
\label{subsec:finding_modes}

For algorithms that do not make explicit use of known mode locations, moving between the modes and finding modes are equivalent tasks. Indeed, a Markov chain with local dynamics that left mode A and moved to mode B will need to rediscover mode A via local moves to return there. As considered in a number of approaches \citep{andricioaei2001smart, lan2014wormhole, MR1844357, zhou2011multi, Pompe2018}, this could be avoided by making use of known mode locations, for example through the jump move discussed in the previous section.

Assume that $\pi$ has $k$ local modes associated with local maxima $\braces{\nu_{1}, \dots, \nu_{k}} \in \mathcal{X}$. In a continuous state space $\mathcal{X}$ with differentiable target $\pi$, identifying modes is equivalent to visiting their basins of attraction (c.f. \citet[Section 2]{zhou2011multi}) in the following sense of gradient ascent. Define the vector field
\begin{equation}
   \psi(x) \coloneq \nabla \pi(x) : \mathcal{X} \to \mathbb{R}^{d},
\end{equation}
which specifies the gradient ascent dynamics for $\pi$ in $\mathcal{X}$. For numerical stability, we may replace $\nabla \pi(x)$ by $\nabla \log(\pi(x))$. If this vector field has a unique integral curve solution 
\begin{equation}
    \Psi(x,t): \mathcal{X} \times \mathbb{R}_{+} \to \mathcal{X},
\end{equation}
then for a starting location $x \in \mathcal{X}$ we can consider the gradient ascent path $\Psi(x,t)$, $t \in [0, \infty)$. 

Define $\Psi(x,\infty) \coloneq \lim_{t \to \infty} \Psi(x,t)$. Under mild regularity conditions this limit will exist and $\Psi(x,t)$ will converge to $\Psi(x,\infty) \in \braces{\nu_{1}, \dots, \nu_{k}}$, that is, one of the maxima of the local modes of $\pi(x)$. We can now define the basins of attraction for each mode as
\begin{equation} 
    \label{def:basin}
    B_{i} = \braces{x \in \mathcal{X}: \Psi(x, \infty) = \nu_{i}}, \qquad i =1, \dots, k.
\end{equation}

\begin{figure}
    \centering
    \includegraphics[scale=.66]{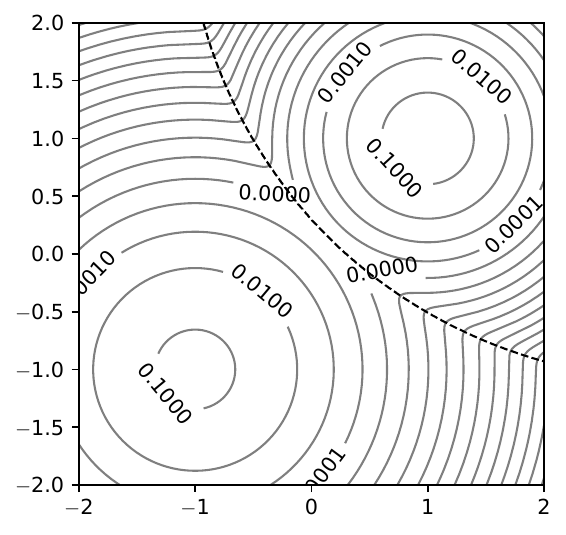}
    \caption{Basins of attraction defined by \eqref{def:basin} for a mixture of two Gaussians with unequal variances. Geometry of basins of attraction gets complex in high dimensions and multiple heterogeneous modes.}
    \label{fig:basins}
\end{figure}

Under some regularity conditions, including that the set of stationary points of $\pi$ has measure zero under $\pi$, $\braces{B_{i}}_{i = 1}^k$ is a partition of $\mathcal{X}$. For the purpose of mode finding, it is enough to obtain a single point from $B_{i}$ and ascend along the gradient to $\nu_{i}$, and establishing further properties of $\braces{B_{i}}_{i = 1}^k$ is not necessary. Note that numerical basins of attraction as defined by a particular optimization procedure may deviate from the ``correct" basins of attraction defined by \eqref{def:basin} (see \citet{asenjo2013visualizing} for a discussion and guidance on optimization algorithms), but since the goal is to find the modes, rather than accurate estimation of basins of attraction, we do not consider this distinction any further. There are various approaches to obtaining points from every basin of attraction $B_{i}$, such as deterministic starting grids over a compact set, or random sampling from a dispersed distribution. The geometry of the basins of attraction becomes complicated as dimension grows, modes are not homogeneous and/or numerous. Figure \ref{fig:basins} illustrates the simple case of a mixture of two Gaussians with unequal variances in $\mathbb{R}^{2}$. Satisfactory theory of how to approach  finding all $B_{i}$'s, when this problem is feasible, or what its complexity is, doesn't seem to be available, maybe because a rigorous formulation of these questions is challenging and case specific. The multimodal algorithms that follow these strategies of finding basins of attraction and building a list of modes before the main MCMC scheme is started, are often referred to as optimization based methods. 

As a side note, should $\nabla \pi(x)$ be not available, expensive to compute, or should gradient ascent be unstable, stochastic gradient methods or simulated annealing \citep{MR0702485,bertsimas1993simulated} can be employed, albeit with a faster cooling schedule to converge to the set of local, rather than global, maxima \citep{MR1377584, MR1700696}. This is particularly relevant in discrete state spaces where there are no gradients, and maxima are ``local" with respect to the adjacency structure of the state space, as defined by the moves of the Markov chain in question.

\subsection{Sampling Efficiently within the Modes}
\label{subsec:sampling_within}

The beauty of the Metropolis-Hastings algorithm lies in its generic construction. Any irreducible proposal kernel $Q$ and an accept-reject step as in \eqref{eq:acc1}, result in an ergodic Markov chain and valid inference in the limit of infinite computational effort. However, in practice MCMC efficiency depends on the choice of $Q$, and a poor choice will result in unreliable inference. The standard procedure is to fix a parametric form for the proposal kernel $Q$, and then tweak its parameters to optimize mixing. This can be done manually, or preferably via an adaptive MCMC design \citep{rosenthal2011optimal}. Similarly to Metropolis-Hastings, other off-the-shelf MCMC samplers, including MALA \citep{MR1440273, MR1625691}, HMC \citep{MR3960671, neal2011mcmc}, Gibbs samplers and Metropolis-within-Gibbs samplers \citep{smith1993bayesian, gelfand1990sampling}, Crank-Nicolson \citep{MR1723510, MR2444507, MR3135540}, or pseudomarginal algorithms \citep{MR2502648, MR3285606, MR3371005}, require tuning of their parameters to produce useful estimates.

The difficulty of multimodal targets in this context comes from heterogeneity between the modes, which may have very different scales and geometry. If the local transition kernel is optimized for one particular mode, it may mix poorly within other modes. In some contexts, the extent to which heterogeneity in scale and geometry affects MCMC efficiency can be assessed explicitly, at least at the theoretical level, using diffusion limits and speed function, as discussed in Section \ref{sec:intro} (see \cite{rosenthal2011optimal} for more details). Following the diffusion limit approach of \citet{MR1428751, MR1888450}, see also \citet{MR2749836}, the efficiency of a Metropolis-Hastings algorithm can be measured through two crucial properties. Firstly, appropriate scaling of the proposal is assessed by comparing the acceptance rate of the algorithm to its optimal rate of $0.234$. Secondly, with respect to geometry, the mismatch between $\Sigma_P$, the covariance matrix of the proposal, and $\Sigma$, the covariance matrix of the target, is measured by the \emph{inhomogeneity factor}. For a $d$-dimensional target distribution, and under certain (admittedly, rather restrictive) regularity conditions, the inhomogeneity factor is defined as
\begin{equation} 
    \label{eq:inhomo}
    b = d \frac{\sum_{i=1}^{d} \lambda_{i}^{-1}}{\parens{\sum_{i=1}^{d} \lambda_{i}^{-1/2}}^{2}},
\end{equation}
where $\braces{\lambda_{i}}_{i=1}^{d}$ are the eigenvalues of $\Sigma^{-1} \Sigma_{P}$. Note that by Jensen's inequality, $b \geq 1$ and $b = 1$ only if $\Sigma_P$ is proportional to $\Sigma$. The value of $b$ is precisely the factor by which convergence is slowed down when using the proposal covariance $\Sigma_P$, relative to the optimal choice of $\Sigma$. While multimodal targets do not satisfy the regularity conditions that underlie this interpretation, the dynamics of the local moves within each mode are sufficiently close to those conditions to draw useful conclusions. Let the local geometry of modes $A$ and $B$ be described by their respective local covariance matrices, denoted $\Sigma_{A}$ and $\Sigma_{B}$. If we optimize the Metropolis-Hastings sampler for mode $A$, i.e. set the proposal covariance matrix $\Sigma_P = \Sigma_{A}$ and then use it within mode $B$, the resulting slowdown is given by \eqref{eq:inhomo}, with $\braces{\lambda_{i}}_{i=1}^{d}$ equal to the eigenvalues of $\Sigma_{B}^{-1}\Sigma_{A}$.

A similar analysis can be carried out if the local sampler within each mode is the random scan Gibbs sampler. There, the optimal coordinate selection probabilities may differ between the modes according to their respective geometries, and the effect of suboptimal selection probabilities on mixing can be assessed \cite{chimisov2018adapting}. We expect that similar inefficiencies arise for other MCMC algorithms, whenever the algorithm is tuned to a particular mode within a collection of heterogeneous modes.

When considering heterogeneity of the modes, and the resulting inefficiency if inadequate parameters are used locally, it becomes evident that separate tuning within each mode would be worthwhile. Ideally, these locally optimal parameters $\gamma \in \mathcal{Y}$ of a $\pi$-invariant transition kernel $P_{\gamma}$ should be learned automatically on the fly, rather than hand-tuned by a human expert in an involved and time consuming process. Learning optimal parameters of a transition kernel on the fly is precisely the idea behind adaptive MCMC algorithms \citep{MR1828504, MR2461882, MR2749836} which generate the process $X_{n}$ by repeating the following two steps:
\begin{enumerate}
    \item Sample $X_{n+1} \sim P_{\gamma_{n}}(X_{n}, \cdot)$; 
    \item Given $\braces{X_{0}, \dots, X_{n+1}, \gamma_{0}, \dots, \gamma_{n}}$, update $\gamma_{n+1}$ according to some adaptation rule.
\end{enumerate}
Here $\gamma_{n} \in \mathcal{Y}$ represents the choice of kernel to be used when updating from $X_{n}$ to $X_{n+1}$. In the adaptation rule above, $\gamma_{n+1}$ is a realization of a random variable $\Gamma_{n+1}$, sampled conditionally on the history $\braces{X_{0}, \dots, X_{n+1}, \gamma_{0}, \dots, \gamma_{n}}$. Adaptive MCMC introduces a theoretical complication in that an algorithm which updates the transition kernel based on its trajectory is not Markovian, so the standard ergodic arguments do not apply. In particular, the distribution of $X_{n}$ started from $x \in \mathcal{X}$ and $\gamma \in \mathcal{Y}$,
\begin{equation}
    D_{n}^{(x,\gamma)}(A) \coloneq \mathbb{P}(X_{n} \in A | X_{0} = x, \Gamma_{0} = \gamma), \qquad A \in \mathcal{B}(\mathcal{X}),
\end{equation}
can no longer be written as \eqref{eq:kernel_iterate}, or even an inhomogeneous product of Markov kernels. It accounts for integrating out not only $\{X_{0}, \dots, X_{n-1}\}$ but also $\braces{\Gamma_{0}, \dots, \Gamma_{n-1}}$. To make sense of ergodicity of Adaptive MCMC, write
\begin{equation}
    T_{n}(x, \gamma) \coloneq \| D_{n}^{(x,\gamma)}(\cdot) - \pi(\cdot)\|_{TV} = \sup_{A \in \mathcal{B}(\mathcal{X})} |D_{n}^{(x,\gamma)}(A) - \pi(A)|
\end{equation}
and call the adaptive algorithm ergodic if 
\begin{equation}
    \label{eqn:adap_erg}
    \lim_{n \to \infty} T_{n}(x, \gamma) = 0 \quad \text{for all} \quad x \in \mathcal{X}, \gamma \in \mathcal{Y}.
\end{equation}
Ergodicity of Adaptive MCMC and other asymptotic validity results have been established under various regularity assumptions on the target density $\pi$, the family of transition kernels in use $\{P_{\gamma}\}_{\gamma \in \mathcal{Y}}$ and the design of the adaptation mechanism in step 2 above. We refer to \citet{MR2340211} for coupling based ergodicity results, to \citet{MR3012408} for martingale approximation techniques, to \citet{chimisov2018air} for computationally efficient strategies that adapt increasingly rarely, and to \citet{Pompe2018} for a general auxiliary variable Adaptive MCMC setting that fits many of the complex MCMC designs used in challenging applications.

Off the shelf adaptive strategies aim to find globally optimal parameters of the transition kernel in question. Regional adaptation, where samplers are designed to locally optimize parameters in each mode, is not straightforward and requires new ideas. It has been explored in several papers \citep{MR2750572, MR2816538, Pompe2018, MR4267920}. The key challenge that these methods face is not only to optimize the parameters, but also to learn an appropriate partitioning of the state space into regions where these parameters should be learned and applied.

\section{Algorithms}
\label{sec:alg}

\subsection{Parallel Tempering}
\label{subsec:alg:temp}

\begin{figure}
    \centering
    \includegraphics[scale=.5]{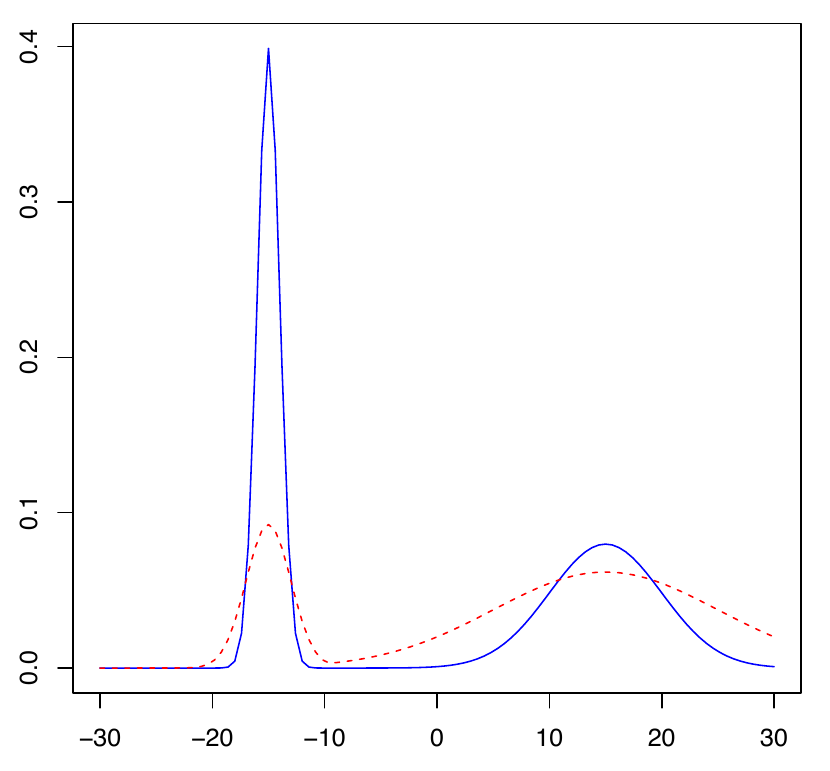}
    \caption{One dimensional multimodal density without tempering (blue) and with tempering (red dashed). Tempered distribution has modes less tall and valleys less deep. Note how tempering reduces the probability mass of the narrow mode.}
    \label{fig:temper}
\end{figure}

The most common and generic approach to dealing with multimodal targets is known as tempering \citep{MR0869788, geyer1991markov, marinari1992simulated}. It makes use of ``tempered" distributions with densities proportional to $\pi^{\beta}$ for some $\beta \in (0,1)$. Having physical systems in mind, the parameter $\beta$ is interpreted as inverse temperature.
The tempered density $\pi^{\beta}$ preserves the location of the modes but is flatter, with the modes being less tall and the valleys less deep, and therefore easier to explore for local transition kernels (see Figure~\ref{fig:temper}).

Parallel tempering (PT) is a canonical way to use tempered distributions that we now describe. In PT one uses a sequence of $L$ inverse temperatures $\boldsymbol{\beta} \coloneq (\beta^{(1)}, \dots, \beta^{(L)})$, such that $1=\beta^{(1)} > \dots > \beta^{(L)} > 0$, and defines a Markov chain $X_{n} = (X_{n}^{(1)}, \dots, X_{n}^{(L)})$ on an augmented space $\mathcal{X}^{L}$, that is, on an $L-$fold Cartesian product of the original state space. The PT chain $X_{n}$ is constructed to be reversible with respect to the target density 
\begin{equation}\label{eqn:temp_target}
    \pi_{\boldsymbol{\beta}}(x^{(1)}, \dots, x^{(L)}) \propto \pi^{\beta^{(1)}}(x^{(1)}) \times \cdots \times \pi^{\beta^{(L)}}(x^{(L)}) \quad \textrm{on} \quad \mathcal{X}^{L}.
\end{equation}
This implies that coordinates $X_{n}^{(l)}, l = 1, \dots, L$, targets $\pi^{\beta^{(l)}}$ and in particular $X_{n}^{(1)}$ targets~$\pi$.

The key idea underpinning the construction is that high temperature chains, in particular $X_{n}^{(L)}$, targeting flattened $\pi$, will mix efficiently between modes and their location should be propagated up the inverse temperature ladder to aid between mode mixing of~$X_{n}^{(1)}$. To this end, PT utilizes two consecutive $\pi_{\boldsymbol{\beta}}-$invariant moves: the propagation move and the swap move. For the propagation move, $X_{n}$ evolves through the application of some type of MCMC transition kernel (such as a random-walk Metropolis) at each temperature independently. That is, for $x, y \in \mathcal{X}^{L}$, the transition kernel of the propagation move is 
\begin{equation}
    \label{eqn:prop_move} 
    M_{\beta}(x,y) = \prod_{l=1}^{L} M_{\beta^{(l)}}(x^{(l)}, y^{(l)}),
\end{equation}
where each $M_{\beta^{(l)}}$ is $\pi^{\beta^{(l)}}$ invariant. The Swap move consists of choosing $l \in \{1, 2, \dots, L-1\}$ uniformly at random to pick two adjacent temperatures, $l$ and $l+1$, and then proposing to swap the current values $x_{n}^{(l)}$ and $x_{n}^{(l+1)}$ of $X_{n}^{(l)}$ and $X_{n}^{(l+1)}$. This proposal is accepted according to the usual symmetric Metropolis acceptance probability targeting $\pi_{\boldsymbol{\beta}}$ of \eqref{eqn:temp_target}, that is,
\begin{equation}
    \min\braces*{1, \frac{\pi^{\beta^{(l+1)}}(x_{n}^{(l)}) \pi^{\beta^{(l)}}(x_{n}^{(l+1)})}{\pi^{\beta^{(l)}}(x_{n}^{(l)}) \pi^{\beta^{(l+1)}}(x_{n}^{(l+1)})}}.
\end{equation}
Let $S_{\beta}^{(l)}$ denote the resulting kernel that attempts to swap process values in temperatures $l$ and $l+1$, then the overall transition kernel of the swap move is 
\begin{equation}
    S_{\beta} = \sum_{l=1}^{L-1} \frac{1}{L-1} S_{\beta}^{(l)},
\end{equation} 
And the combined transition kernel of the PT is $P_{\beta} = M_{\beta} S_{\beta}$.

Although PT is a generic approach, there are some design decisions to be made when implementing it. Too few temperature levels will result in either (i) the hottest temperature being not hot enough and the hottest chain still not mixing between the modes; or (ii) the gaps between temperature being too wide and the swap moves not being accepted. With too many temperature levels, not only does the implementation cost grow, but also, due to the diffusive behavior of the swap process, the states from the hot temperatures need $O(L^{2})$ swap steps to travel up the ladder and influence the cold chains, which results in slower mixing \citep{MR2826692}.  Hence, tuning the number of temperatures and the temperature spacing, as well as the parameters of the propagation kernels $M_{\beta^{(l)}}$, is essential in more complex problems. To facilitate this optimization, \cite{MR2826692} provided an analogue of the random walk Metropolis optimal scaling result and showed that the inverse temperatures should be spaced so that the probability of accepting a temperature swap is approximately 0.234. Optimal scaling results for PT and RWM lead to the adaptive PT version of \cite{Miasojedow2013}, which adapts both the temperature spacing and the propagation kernels. The temperature spacing is adapted by adjusting the gaps between inverse temperatures to target 0.234 for each pair, while propagation kernels follow the well established Adaptive Random Walk Metropolis design \citep{MR1828504} learning proposal scales and covariance matrices at each temperature levels. To reduce the number of parameters in the adaptation, a common covariance matrix can be used at all temperatures. Separate scaling for each of the temperatures will then account for the necessary modifications of the proposals. The refined adaptive PT version of \cite{lkacki2016state} also adapts the number of temperatures. Both versions enjoy asymptotic ergodicity in the sense of \eqref{eqn:adap_erg} for multimodal targets with exponentially or faster decaying tails satisfying some natural regularity conditions on curvature.

An interesting aspect of PT is that the propagation moves are independent at each temperature which motivates parallel or distributed implementations \citep{altekar2004parallel, mingas2012parallel, fang2014parallel, syed2022non}. For such implementations it is sensible to assume that the cost of PT per iteration is independent of $L$, the number of temperatures. The remaining problem that the information transfer from hot to cold temperatures takes $O(L^{2})$ time due to the diffusive behavior has been recently resolved in \cite{syed2022non}. For this approach, consider the maximum collection of swap moves that can proposed in parallel without interference. There are two such choices:
\begin{equation}
    S_{\beta}^{\mathrm{even}} \coloneq \prod_{\text{$l$ even}}S_{\beta}^{(l)}, \qquad S_{\beta}^{\mathrm{odd}} \coloneq \prod_{\text{$l$ odd}} S_{\beta}^{(l)}.
\end{equation}
For a parallel implementation of PT, one needs to make a choice how to alternate between $S_{\beta}^{\mathrm{even}}$ and $S_{\beta}^{\mathrm{odd}}$. Choosing uniformly at random results in the swap move kernel becoming $S^{\text{R}}_{\beta} = \frac{1}{2}S_{\beta}^{\mathrm{even}} + \frac{1}{2}S_{\beta}^{\mathrm{odd}}$ and a reversible algorithm, however it does not cure the diffusive behavior. Alternating sequentially between $S_{\beta}^{\mathrm{even}}$ and $S_{\beta}^{\mathrm{odd}}$ results in 
\begin{equation}
    S^{\text{NR}}_{\beta} = \begin{cases}
        S_{\beta}^{\mathrm{even}}, & \text{if $n$ is even}, \\
        S_{\beta}^{\mathrm{odd}}, & \text{if $n$ is odd};
    \end{cases}
\end{equation}
as considered by \cite{syed2022non} (see also \cite{okabe2001replica, lingenheil2009efficiency}), and a nonreversible PT algorithm which, if tuned correctly, can transfer the information from hot to cold temperatures in $O(L)$ steps.

Parallel Tempering enjoys great popularity due to the appealing intuition that underpins it, negligible implementation overhead of the swap move, and due to the well developed adaptive and parallel versions. However, PT has some shortcomings: (i) it does not remember the visited regions and therefore jumping between the modes is as difficult as finding them; (ii) it does not provide a solution to the necessity of local sampling of modes with different geometry.

One of the key questions regarding PT is how its mixing scales with dimension. There are two very different scenarios to consider. The positive story is that for regular multimodal distributions that have the same width of modes and for which the cold chains mix rapidly within the modes and the hot chains mix rapidly between the modes, these properties carry over to PT and it mixes rapidly as well \citep{woodard2009sufficient}. Intuitively, this is because starting from one mode, the easiest path to another one is to go down the temperature ladder using swap moves, then access the other mode in high temperature level and go up the temperature ladder via swap moves again. Therefore, under these regularity assumptions, no cut will exist separating these two modes with exponentially decaying terms in \eqref{eq:conductance_local}. The rigorous proof builds on a beautiful decomposition argument of \cite{MR1910641} and \cite{MR1943860}. The second setting is that of modes of different width, where the spikiness of one mode compared to another becomes more severe with dimension, as in Example \ref{subsec:mix}. For this type of target distributions the spiky mode looses probability mass in hot temperatures. This can already be seen in one dimension, note how in Figure~\ref{fig:temper} heating the distribution reduced the probability mass of the left narrow mode. As a consequence of this phenomenon, the hot chains do not mix between the modes because the narrow mode has an exponentially decaying probability mass. Since the hot chains do not know about the narrow modes, they can not transfer the relevant information along the temperature ladder to the cold chains. The path from one mode to the other that was feasible in the symmetric case now has a bottleneck and a cut similar to \eqref{eq:conductance_local} can be constructed, implying that the PT chain is torpidly mixing and convergence rates decay exponentially with dimension \citep{woodard2009sufficient}.

\subsection{Wang--Landau}
\label{subsec:wang}

The algorithm of \citet{Wang2001a} was originally proposed for sampling from discrete distributions that arise in statistical physics, such as the Bolzmann machine or the Ising model. This algorithm was adapted for general state spaces by \citet{liang2005generalized} and \citet{atchade2010wang}. The idea is to select a low-dimensional function of the parameters of the model $\xi(x)$, known as a reaction coordinate. The most typical choice is the unnormalized negative log-density, so that $\xi(x)$ is equal to $-\log \pi(x)$ up to an additive constant. \citet{chopin2012free} provide a detailed comparison of different choices of $\xi(x)$ for a univariate Gaussian mixture model. For the purpose of simplification, in the following we assume that $\xi(x)$ is one-dimensional.

The Wang-Landau algorithm targets a biased distribution $\tilde{\pi}^{\xi}(x)$, such that $\xi(x)$ is uniformly distributed over the interval $[z_{\mathrm{min}}, z_{\mathrm{max}}]$. The purpose of biasing the target in this way is to add more weight to the low probability regions and make valleys less deep. Ideally, $\tilde{\pi}^{\xi}(dx \mid \xi(x) = z)$ should be less multimodal than the original target $\pi(x)$, so that the algorithm can explore the state space more efficiently, speeding up convergence. One of the difficulties in applying this algorithm in general state spaces is choosing suitable bounds $z_{\mathrm{min}}$ and $z_{\mathrm{max}}$, since for sensible choices of $\xi(x)$ the biased target $\tilde\pi^\xi(x)$ will not be integrable on an unbounded domain. The state space $\mathcal{X}$ is partitioned into $J$ regions $\braces{\mathcal{B}_{j}}_{j=1}^{J}$ according to $\xi(x)$, meaning that $\xi(x)$ is increasing in $j$. At each epoch $t=1,\dots,T$, MCMC samples $\braces{X_{n}}_{n=1}^{N}$ are obtained using a biased transition kernel $Q(x,\cdot)$, where the bias $\gamma_t(j)$ pushes the chain towards regions that have been visited less.

Let $\nu_{t}(j)$ be the proportion of time spent by $\braces{X_{n}}$ with $\xi(X_{n}) \in \mathcal{B}_{j}$ during the current epoch, then all $\nu_{t}(j) \rightarrow J^{-1}$ as $n \rightarrow \infty$, this is known as the flat histogram criterion. Once this criterion is approximately satisfied, the bias is updated adaptively and all of the relative occupancy rates $\nu_{t+1}(j)$ are reset to 0 for the next epoch. After $T$ epochs have been run, the samples then need to be post-processed to correct for the bias, for example using importance sampling \citep{chopin2012free} or sequential Monte Carlo \citep{Chopin2010smc}.

\citet{bornn2013adaptive} introduced a parallel, adaptive Wang--Landau algorithm (PAWL) that runs $M$ interacting chains in parallel. As well as updating the bias $\gamma_t(j)$ and the MCMC kernel $Q$ adaptively using information from all $M$ chains, the algorithm can also split the partitions $\mathcal{B}_{j}$, adaptively increasing the total number of histogram bins as needed. We demonstrate PAWL with a mixture of Gaussians in Section~\ref{subsec:mix} and with an Ising model in Section~\ref{subsec:ising}. Although the algorithm works well for the Ising model, which has a discrete state space, we find that it scales very poorly in dimension for the mixture model. This is due to the need to set $z_{\mathrm{max}}$ very far out into the tails to overcome the problems discussed in Section~\ref{subsec:between_modes} and illustrated by Figure~\ref{fig:acc-rej}.

\subsection{Mode Jumping and Regional Adaptation Approaches}
\label{subsec:jump}

The concept of mode jumping was first introduced by \citet{MR1844357}. Their original algorithm involves a mixture of two proposal distributions, $Q_{1}$ and $Q_{2}$. $Q_{1}$ can be any Markov kernel that uses local moves to mix within the modes, as discussed in Section~\ref{subsec:sampling_within}. $Q_{2}$ uses a Gaussian random walk proposal with a very large variance, big enough to jump from one basin of attraction into another. Usually the acceptance probability of such a proposal would be vanishingly small, but the mode-jumping algorithm combines MCMC with gradient-based optimization to find the nearest local maximum $\nu_{n+1}$ to the sampled point $z_{n+1}$. The Hessian matrix of second-order partial derivatives is used to construct a local Laplace approximation, $\mathcal{N}\left(\nu_{n+1}, (-H_{n+1})^{-1} \right)$, and a new proposal $y$ is sampled from this distribution. In order to preserve reversibility \eqref{eq:rev}, it is also necessary to calculate the probability of jumping backwards $q_{2}(X_{n} \mid y)$. Let $\delta_{n+1} = z_{n+1} - X_{n}$ be the vector from $X_{n}$ to the sampled point $z_{n+1}$, then $z_{-n} = y - \delta_{n+1}$ represents moving in the opposite direction, away from $\nu_{n+1}$. It is therefore likely that this point $z_{-n}$ lies within the basin of attraction of a different mode. A second optimization is performed to find this local maximum, $\nu_{-n}$, and construct another Laplace approximation to evaluate $\mathcal{N}(X_{n}; \nu_{-n}, (-H_{-n})^{-1})$. This procedure is illustrated by Figure~1 in \citet{MR1844357}.

Although mode-jumping proposals can improve mixing in some circumstances, the accuracy of the Laplace approximation degenerates exponentially quickly as the dimension increases \citep{liu1996metropolized}. This is characterized by the spectral gap in \eqref{eq:jump_gap}. The computational cost of performing two optimizations per iteration of MCMC can also be quite high. The original algorithm is memoryless, so these optimizations might find same local optima $\nu_{n+1}$ and $\nu_{-n}$ over and over again, although \citet{Tjelmeland2004} proposed an adaptive version of the algorithm. There are several related approaches, including darting Monte Carlo \citep{andricioaei2001smart, sminchisescu2011generalized, ahn2013distributed} and wormhole HMC \citep{lan2014wormhole}. Below we describe two in particular, the Jumping Adaptive Multimodal Sampler \citep[JAMS;][]{Pompe2018} and the Annealed Leap-Point Sampler \citep[ALPS;][]{Roberts2020,tawn2021alps}.

\subsubsection{JAMS - Mode Jumping within an Adaptive Framework}
\label{subsec:jams}

In contrast to the mode jumping approach introduced above, the JAMS algorithm \citep{Pompe2018} consolidates the tasks of mode discovery and local geometry estimation to preliminary steps, obviating the need for repeated optimizations at mode jumps. The main phase then consists of running a Markov chain on the augmented posterior density
\begin{equation}
    \label{eqn:JAMS_posterior}
    \pi(x, i) = \pi(x) \frac{w_{i} \kappa_{i}(x)}{\sum_{j=1}^{\mathcal{I}} w_{j} \kappa_{j}(x)},
\end{equation}
where $i$ indexes modes, $w_{i}$ is the weight of the $i$-th component, and $\mathcal{I}$ is the index set of modes. $\kappa_{i}(x)$ is typically specified as a Gaussian or Student-t distribution centered on mode $i$, with some covariance matrix that captures the local geometry. This connects the notions of local and global moves to Gibbs sampling, with updates to $x$ corresponding to local moves, and $i$ to global moves. Since the algorithm keeps track of which mode it is currently targeting, such samples can be used to estimate the posterior covariance conditional on $i$, which as an estimate of local geometry is more robust than a Laplace approximation. Such estimates enable mode-specific proposals for local moves, as motivated in Section \ref{subsec:sampling_within}, as well as efficient jump kernels based on proposals that account for difference in geometries.

To initialize the augmented density, two preliminary phases are carried out. In the first phase, local modes are discovered by running an optimization routine from a collection of starting values. It is sufficient for those starting values to populate all the basins of attraction, as set out in Section \ref{subsec:finding_modes}. Once the location of the modes is known, their geometry is estimated in a second phase, where separate Markov chains target the various conditionals $\pi(x|i)$. While transition criteria between phases are necessarily ad-hoc and rely on heuristics, the first phase may be continued in parallel to the main run, wherein any local geometry estimate can also be further refined through standard adaptive MCMC strategies. For the theoretical underpinning of JAMS adaptive MCMC theory is extended to augmented state space setting that accommodates \eqref{eqn:JAMS_posterior}.

\subsubsection{ALPS}
\label{subsec:alps}

The key feature that distinguishes ALPS from other mode-jumping algorithms is the use of regional weight-preserving tempering \citep{GarethJeffNick} for \emph{cold} temperatures; that is for tempered targets $\pi(x)^\tau$ with powers $\tau > 1$. This is to address the problem with degeneracy of Laplace approximations in high dimensions, discussed in Section~\ref{subsec:between_modes}. Even if the local region surrounding a mode is poorly approximated by a Gaussian at the target temperature $\tau = 1$, for example due to skewness or heavy tails, the approximation error can be reduced by increasing $\tau$. Note that preserving the relative probability mass surrounding each mode is essential here, or else all of the mass would concentrate around a single mode, as in simulated annealing \citep{bertsimas1993simulated}.

ALPS uses parallel tempering with an increasing sequence $\tau_{0} = 1 < \dots < \tau_{T}$. Mode-jumping is only performed at the coldest temperature $\tau_{T}$, where the acceptance probability is high. This avoids the usual problem with torpid mixing of parallel tempering \citep{woodard2009sufficient}. Between-temperature swap moves are accelerated using a deterministic transformation \citep{Tawn2018}, while within-temperature moves use preconditioned RWM proposals. The exploration component of ALPS combines RWM at a hot temperature $0 < \tau_{\mathrm{EC}} < 1$ with gradient ascent to search for the modes. This enables it to move between basins of attraction and assemble a list of known modes with corresponding Laplace approximations for mode-jumping proposals. ALPS therefore addresses all of the fundamental challenges described in Section~\ref{sec:challenges}.

\subsection{Repelling-Attracting Ideas}
\label{sec:ra}

Repelling-attracting algorithms are characterized by a particular approach to constructing the Metropolis-Hastings proposal kernel $Q$, usually in continuous state spaces. The kernel generates a proposal in two steps, the first of which is biased towards areas of lower density (repelling phase), while in the second it moves back towards a higher density area (attracting phase). Ideally, the repelling moves into the low-density barrier between 2 modes, while the attracting move recovers enough density to result in acceptance of the proposal under the usual Metropolis-Hastings rule.

An earlier exponent of this principle is \emph{Repelling-attracting Metropolis} (RAM), introduced in \citet{Tak2016}. It starts by forcing an intermediate downward move by repeatedly proposing according to a symmetric random walk kernel $R(x, \cdot)$, and accepting with probability
\begin{equation}
    \alpha^{\downarrow}(x, z) = \min\braces*{1, \frac{\pi(x)}{\pi(z)}}
\end{equation}
until the first acceptance. This is followed by repeatedly proposing according to $R(z, \cdot)$, and accepting with probability
\begin{equation}
    \alpha^{\uparrow}(z, y) = \min \braces*{1, \frac{\pi(y)}{\pi(z)}}
\end{equation}
until the first acceptance, thereby forcing an upward move. The composition of the two steps yields the proposal kernel $Q$. Since the density corresponding to $Q$ is itself intractable, the authors introduce an additional auxiliary variable, which results in a tractable joint acceptance probability. This puts RAM in the category of pseudomarginal algorithms. In practice, successful use of RAM requires careful hand-tuning of $R$, since no appropriate adaptive rule has been proposed.

A more recent variant is \emph{Repelling-attracting Hamiltonian Monte Carlo} (RAHMC), introduced by \cite{Rep_Att_HMC}, which replaces the random walk-based proposal mechanism with a suitable modification of the Hamiltonian dynamics that are at the heart of the default HMC algorithm. Hamiltonian dynamics approximately preserve the augmented target
\begin{equation}
    \pi(\boldsymbol{x}, \boldsymbol{z}) \propto \pi(\boldsymbol{x}) \exp\braces*{-\boldsymbol{z}^{\top} \Sigma^{-1} \boldsymbol{z} / 2},
\end{equation}
where $\boldsymbol{z}$ is a vector of dimension equal to $\boldsymbol{x}$, and $\Sigma$ is a positive definite matrix. This preservation of the augmented target makes mode jumping in the marginal $\pi(\boldsymbol{x})$ very unlikely. In analogy to RAM, the proposed adaptation consists of modifying the Hamiltonian dynamics in 2 phases, with $\pi(\boldsymbol{x}, \boldsymbol{z})$ decreasing in a first phase of dissipating dynamics, and then increasing again in a second phase of amplifying dynamics. If the dissipating and amplifying phases are sufficiently long, the combined RAH dynamics may jump to a new mode. To this end, \cite{Rep_Att_HMC} propose a tuning scheme for the parameters that control the intensity of dissipation and amplification, and the length of the phases. Moreover, \cite{Rep_Att_HMC} show that the RAH dynamics preserve various important properties of Hamiltonian dynamics - in particular, while $\pi(\boldsymbol{x}, \boldsymbol{z})$ is not necessarily conserved, the drift is observed to be small, resulting in a high acceptance probability.

\section{Examples}
\label{sec:examples}

This section contains a synthetic example and two real-world examples of multimodal distributions. We use the synthetic example of a mixture of Gaussians to show how various algorithms from Section~\ref{sec:alg} perform with increasing dimension. The autologistic (Ising) model in Section~\ref{subsec:ising} is an example of a discrete distribution, while the seemingly-unrelated regression (SUR) model in Section~\ref{subsec:sur} is an example of a curved exponential family \citep{Sundberg2010}.

\subsection{Mixture Model}
\label{subsec:mix}

A mixture of two Gaussians with different covariances is a popular numerical example for benchmarking multimodal algorithms, as it captures all three fundamental challenges of multimodal sampling that we describe in Section~\ref{sec:challenges}. Firstly, with increasing dimension, the valley between the modes deepens, and moving between the modes becomes more challenging \citep{woodard2009conditions, woodard2009sufficient}. Similarly, as the relative difference in scale between the modes increases, designing jump moves directly into the other mode becomes harder. Secondly, finding the basin of attraction of the narrower mode is more difficult as dimension increases because its probability mass becomes concentrated on an exponentially small region. Finally, increasing dimension also affects sampling efficiency within each mode, because learning the respective local covariance matrices becomes more difficult.

While this example appears very stylized, it mimics certain multimodal scenarios that may appear in practice. The Bernstein-von Mises theorem \citep{MR1652247} states that, under suitable regularity conditions, the posterior distribution converges in the limit of infinite data to a multivariate normal distribution centered at the maximum likelihood estimator with covariance given by the Fisher information matrix, i.e. $\mathcal{N}(\hat{\theta}_{\mathrm{ML}}, \mathcal{I}(\hat{\theta}_{\mathrm{ML}}))$. In case of model misspecification, the limiting distribution becomes $\mathcal{N}(\hat{\theta}_{\mathrm{KL}}, \mathcal{I}(\hat{\theta}_{\mathrm{KL}}))$, where $\hat{\theta}_{\mathrm{KL}}$ is the unique parameter value $\theta$ minimizing the Kullback-Leibler divergence $\operatorname{KL}(\theta)$ between the unknown true data distribution and the data model $p(\theta | \mathrm{data})$. For finite data, and when the $\operatorname{KL}(\theta)$ has local minima, a mixture of Gaussians with centers at these local minima $\hat{\theta}_{\mathrm{KL}}$ and locally-evaluated Fisher information matrices $\mathcal{I}(\hat{\theta}_{\mathrm{KL}})$ as covariance matrices is a useful approximation of the resulting posterior. Since all models are wrong and data is always finite, studying a mixture of two Gaussians with different covariance matrices is also statistically well motivated!

We follow \citet{Pompe2018} in using an equally-weighted mixture of two multivariate Gaussians:
\begin{equation*}
    \pi(\boldsymbol{x}) = \frac{1}{2} \mathcal{N}\parens{\boldsymbol{x}; \boldsymbol{m}_{1}, \sigma_{1}^{2} \mathrm{I}_{d}} + \frac{1}{2} \mathcal{N}\parens{\boldsymbol{x}; \boldsymbol{m}_{2}, \sigma_{2}^{2} \mathrm{I}_{d}}
\end{equation*}
with increasing dimensions $d \in \braces{2, 4, 8, 16, 32, 64}$, where $\boldsymbol{x} \in \mathbb{R}^{d}$, $\boldsymbol{m}_{1} = -(1,\dots,1)^{\top}$, $\boldsymbol{m}_{2} = (1,\dots,1)^{\top}$, $\sigma_{1}^{2} = 0.5 \sqrt{d/100}$, $\sigma_{2}^{2} = \sqrt{d/100}$, and $\mathrm{I}_{d}$ is the $d \times d$ identity matrix. Since $\sigma_{1}^{2} \ne \sigma_{2}^{2}$, this is sufficient to result in torpid mixing of parallel tempering \citep{woodard2009sufficient}.   

Almost all of the algorithms that we discuss in Section~\ref{sec:alg} are applicable in this setting. In the following, we have selected four algorithms to compare: RWM; adaptive parallel tempering (APT; Section~\ref{subsec:alg:temp}); PAWL (Section~\ref{subsec:wang}); and JAMS (Section~\ref{subsec:jams}). The RWM sampler that we use is implemented in the R package \texttt{NIMBLE} \citep{nimble2017}. It uses a joint, Gaussian proposal distribution $Q(\boldsymbol{x}, \cdot)$ that is automatically tuned using the adaptation scheme of \citet{shaby2010exploring}. We use the APT algorithm of \citet{Miasojedow2013,lkacki2016state}, as implemented in \texttt{nimbleAPT} \citep{nimbleAPT}. The PAWL algorithm of \citet{bornn2013adaptive} is implemented in the R package of the same name \citep{PAWL}. The histogram bins for PAWL are determined using the negative log-density, $\xi(\boldsymbol{x}) = -\log \pi(\boldsymbol{x})$. We have created our own open-source implementation of JAMS \citep{Pompe2018} that is available at \url{https://github.com/timsf/jams}.

\begin{figure} 
    \centering
    \includegraphics[width=0.8\textwidth]{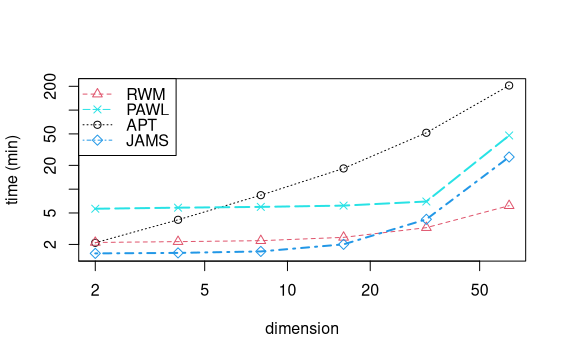}
    \caption{Elapsed wall-clock times (in minutes) for RWM (dashed line with triangles); PAWL (dashed line with crosses); APT (dotted line with circles); and JAMS (dot-dashed line with diamonds) for the mixture model example in Section~\ref{subsec:mix}, with increasing dimension from $d=2$ up to $d=64$. Both axes are on a logarithmic scale.}\label{fig:mixElapsed}
\end{figure}  

The elapsed runtimes of the four algorithms are roughly comparable at $d=2$, as shown in Figure~\ref{fig:mixElapsed}, but their scalability with increasing dimension varies considerably. APT ends up being the most computationally intensive, taking 51.7 minutes for $d=32$ and 204.9 minutes for $d=64$. We run APT for 150,000 iterations with 5 temperature levels. The elapsed times for PAWL includes both the pilot chain, which we run for 50,000 iterations, as well as 500,000 iterations of PAWL itself. We initialize PAWL with $J=10$ histogram bins, but the adaptive algorithm creates more as needed. The runtime of PAWL hardly increases at all from $d=2$ up to $d=32$, due to being able to utilize all of the available CPU cores for parallel computation. However, the cost jumps sharply from 7.0 minutes for $d=32$ to 47.8 minutes for $d=64$. JAMS takes 4.2 minutes for $d=32$ and 25.5 minutes for $d=64$. As expected, the computational cost of RWM scales most efficiently with dimension, since it is the least complicated of the four algorithms. It takes 3.3 minutes for $d=32$ and 6.2 minutes for $d=64$. This is not the complete picture, however, as the abilities of these algorithms to sample from the target distribution do not scale equally well as the dimension increases.

\begin{figure}
    \centering
    \includegraphics[height=3in]{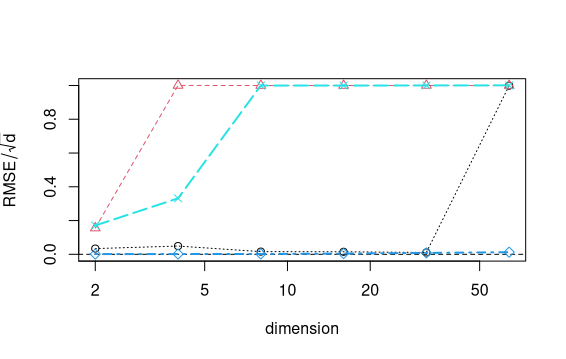}
    \caption{Root Mean Square Error ($\mathrm{RMSE}/\sqrt{d}$) for RWM (dashed line with triangles); PAWL (dashed line with crosses); APT (dotted line with circles); and JAMS (dot-dashed line with diamonds) for the mixture model example in Section~\ref{subsec:mix}, with increasing dimension from $d=2$ up to $d=64$. The horizontal axis is on a logarithmic scale.}
    \label{fig:mixRMSE}
\end{figure}

To quantify the difference between the expectation $\mathbb{E}_\pi[\boldsymbol{x}]$ and its Monte Carlo approximation \eqref{eq:erg_av}, we calculate the Root Mean Square Error (RMSE) divided by the square root of the dimension of $\mathcal{X}$. Since the two mixture components are evenly weighted, we know for this example that $\mathbb{E}_\pi[\boldsymbol{x}] = \boldsymbol{0}$. The RMSE is then the Euclidean distance of the ergodic average of the MCMC samples $\braces{X_{n}}_{n=0}^{m}$ from the origin. The results for all four algorithms with increasing dimension from $d=2$ up to $d=64$ are shown in Figure~\ref{fig:mixRMSE}. JAMS performs the best out of the four algorithms, with RMSE very close to 0 even in 64 dimensions. When $\mathrm{RMSE}/\sqrt{d}$ approaches 1, this indicates that the algorithm has only sampled from one of the two modes. The performance of APT is very close to JAMS up to $d=32$, but then sharply declines in 64 dimensions. See \citet{Pompe2018} for further comparisons of RMSE between APT and JAMS. 

The performance of RWM and PAWL is much worse, even in $d=2$ where $\mathrm{RMSE}/\sqrt{d}$ is 0.171 for PAWL and 0.156 for RWM. Although Wang-Landau is one of the few algorithms that is applicable for discrete distributions, as illustrated in the following section, its weakness for unbounded, continuous state spaces is that it is reliant on the pilot RWM chain to determine the range of energies $[z_{\mathrm{min}}, z_{\mathrm{max}}]$ to use for the flat histogram criterion. If that pilot chain is not mixing well, as in this example, then the range of energy values will not be chosen correctly. This is where algorithms that combine mode-jumping proposals with Wang-Landau, such as the multi-domain sampler \citep{zhou2011multi}, can be beneficial. The performance of RWM and PAWL worsens as the dimension increases, with $\mathrm{RMSE}/\sqrt{d} \approx 1$ for RWM in $d \ge 4$ and for PAWL in $d \ge 8$. In contrast, we demonstrate in the following section that PAWL performs well even in $d=1600$ for a discrete distribution.

\subsection{Autologistic Model}
\label{subsec:ising}

The autologistic model \citep{besag1974spatial}, also known as the \citet{ising1925beitrag} model in statistical physics, is a type of spatial mixture model that is often used in image analysis:
\begin{equation}
    \label{eq:ising}
    \pi(\boldsymbol{x} \mid \boldsymbol{y}, \alpha, \beta) = \frac{\exp\braces*{\alpha \sum_{i} \delta(x_{i}, y_{i})} \exp\braces*{\beta \sum_{i \sim j} \delta(x_{i}, x_{j})}}{\mathcal{Z}(\boldsymbol{y}, \alpha, \beta)}
\end{equation}
where $\boldsymbol{x}, \boldsymbol{y} \in \braces{0,1}^{d}$ are binary vectors, $\alpha, \beta \in \mathbb{R}_+$ are parameters, $\delta(a,b)$ is the Kronecker delta function, $i \sim j$ is a neighborhood relation between pixels, and $\mathcal{Z}(\boldsymbol{y}, \alpha, \beta)$ is a normalizing constant that involves a sum over all $2^{d}$ possible values of $\boldsymbol{x}$. The sufficient statistics of this model are $s_{1}(\boldsymbol{x}, \boldsymbol{y}) = \sum_{i} \delta(x_{i}, y_{i})$, the count of pixels in $\boldsymbol{x}$ that are identical to $\boldsymbol{y}$; and $s_{2}(\boldsymbol{x}) = \sum_{i \sim j} \delta(x_{i}, x_{j})$, the count of identical neighbors. The neighborhood relation includes pixels in the vertical, horizontal, and diagonal directions, so each pixel can have up to 8 neighbors. Pixels on the boundary of the image domain will have less than 8.

We use a satellite image of ice floes that was originally published in \citet{banfield1992ice}. This data is available in the R package \texttt{PAWL} \citep{PAWL}. The dimensions of the image are $40 \times 40$ pixels, so $d = 1600$. The parameters are fixed at $\alpha = 1$ and $\beta = 0.7$, as in \citet{bornn2013adaptive}. See \citet{vu2023warped} for discussion of posterior inference for $\alpha$ and $\beta$ with the intractable likelihood \eqref{eq:ising}. Here we are focused on inference for the latent states $\boldsymbol{x}$, so the normalizing constant $\mathcal{Z}(\boldsymbol{y}, \alpha, \beta)$ cancels out in the acceptance ratio \eqref{eq:acc1}. This means that it is possible to sample from \eqref{eq:ising} using Gibbs or RWM, but the chains will mix very slowly and will be prone to getting stuck in local modes \citep{higdon1998auxiliary,roberts1997updating}. Most algorithms that rely on mode-jumping proposals, discussed in Section~\ref{subsec:jump}, cannot be applied to this model because the state space is discrete and hence \eqref{eq:ising} is not differentiable with respect to Lebesgue measure.


\begin{figure}
    \centering
    \includegraphics[height=3in]{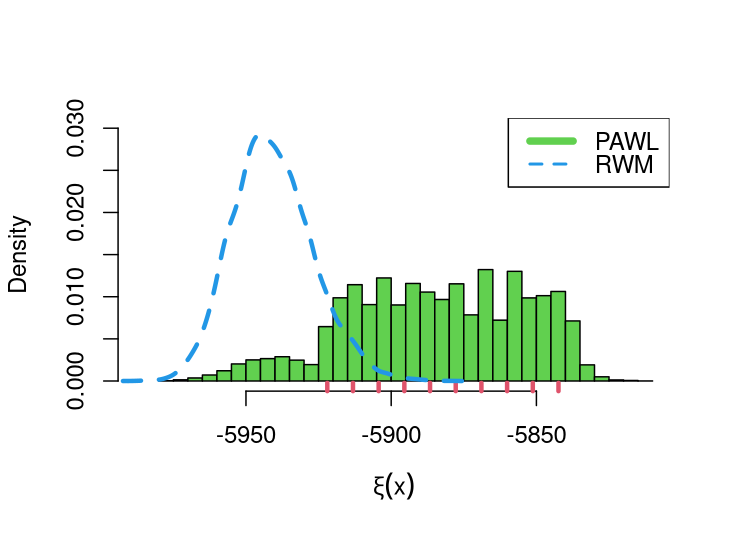} 
    \caption{Histogram of samples from the autologistic model (Section~\ref{subsec:ising}) obtained using the parallel, adaptive Wang-Landau (PAWL) algorithm, in comparison to random-walk Metropolis (RWM).}
    \label{fig:isingHist}
\end{figure}  

We run the PAWL algorithm from Section~\ref{subsec:wang}, determining the histogram bins using the unnormalized log-energy, $\xi(\boldsymbol{x}) = -\alpha s_{1}(\boldsymbol{x}, \boldsymbol{y}) - \beta s_{2}(\boldsymbol{x})$, which is equal to $-\log \pi(\boldsymbol{x} \mid \boldsymbol{y}, \alpha, \beta)$ up to an additive constant. First, a preliminary RWM chain is run for 20,000 iterations to determine the initial range of energy values from $z_{\mathrm{min}}=-5910.4$ to $z_{\mathrm{max}}=-5819.2$, which are evenly divided into $J=10$ bins. It then took 6.5 minutes to run PAWL for 10$^6$ iterations with $M=10$ parallel chains. In comparison, 10$^6$ iterations of RWM with 4 parallel chains took just under 4 minutes. This is consistent with the relative computational cost of RWM and PAWL shown in Figure~\ref{fig:mixElapsed} in the previous section. After a burn-in of 200,000 iterations, RWM converges to a mode centered at -5940. In contrast, PAWL is able to explore a much wider range of energy states, from -5977 to -5811. A histogram of the PAWL samples is shown in Figure~\ref{fig:isingHist}, with the outline of the density plot from RWM shown for comparison.

\subsection{Curved Exponential Family}
\label{subsec:sur}

It is well known that exponential families of models have log-concave likelihood functions and hence a unique, global maximum. Under the canonical parameterization, the dimension of the parameter space is equal to the dimension of the sufficient statistics. However, there is a class of models known as the curved exponential family, for which this is not the case. Often this arises when constraining those parameters, which reduces the effective dimension of the parameter space. Important examples include the Behrens-Fisher model \citep{Sundberg2010}, the seemingly-unrelated regression (SUR) model \citep{Drton2004}, and the exponential random graph model (ERGM), see \citet{chatterjee2013estimating}.

The SUR model is a type of multi-response model \citep[Chapter 14]{Searle2017}. It involves a system of $M$ linear regression equations:
\begin{equation*}
    \boldsymbol{y}_{m} = X_{m} \boldsymbol{\beta}_{m} + \boldsymbol{\epsilon}_{m} ,
\end{equation*}
for $m=1,\dots,M$, where $\boldsymbol{y}_{m}$ is a vector of $n_{m}$ observed responses, $X_{m}$ is an $n_{m} \times J$ matrix of covariates, $\boldsymbol{\beta}_{m}$ is a vector of $J$ unknown regression coefficients, and $\boldsymbol{\epsilon}_{m}$ is a vector of errors. 
The difficulty in estimating $\boldsymbol{\beta}_{m}$ arises because these errors are correlated with a Kronecker product structure,
\begin{equation*}
    \boldsymbol{\epsilon} \sim \mathcal{N}\parens*{ \boldsymbol{0}, \Sigma \otimes \mathrm{I}_{n_{m}}},
\end{equation*}
where $\boldsymbol{\epsilon} = (\boldsymbol{\epsilon}_{1}, \dots, \boldsymbol{\epsilon}_{M n_{m}})^{\top}$, $\Sigma$ is an unknown $M \times M$ variance-covariance matrix, and $\mathrm{I}_{n_{m}}$ is an $n_{m} \times n_{m}$ identity matrix. For example, when $M=2$ and $J=1$, the parameter space $\mathcal{X} \subset \mathbb{R}^5$, constrained so that $\Sigma$ is positive-definite.

\citet{Zellner1962} introduced an iterative, generalized least-squares algorithm for fitting the SUR model. Since then, this model has been widely adopted for panel data in econometrics \citep{Srivastava1987,Fiebig2001}. Convergence of \cite{Zellner1962} algorithm is only guaranteed under the assumption of log-concavity \citep[p. 686]{Greene1997}. However, \citet{Drton2004} showed that the profile log-likelihood $\ell(\boldsymbol{\beta})$ can have up to 5 stationary points and hence $\le 3$ modes in the bivariate case (i.e. $M=2$ and $J=1$). This is because finding the stationary points of $\ell(\boldsymbol{\beta})$ is equivalent to finding the real roots of a fifth-degree polynomial. Given \cite{Zellner1962} estimator $\widehat\Sigma$, the profile log-likelihood is given by
\begin{equation*}
    \ell(\boldsymbol{\beta}) = -n\log\braces{2\pi} - \frac{n}{2} \log |\widehat{\Sigma}| - n.
\end{equation*}

\begin{figure}
\begin{center} 
    \includegraphics[height=2.5in]{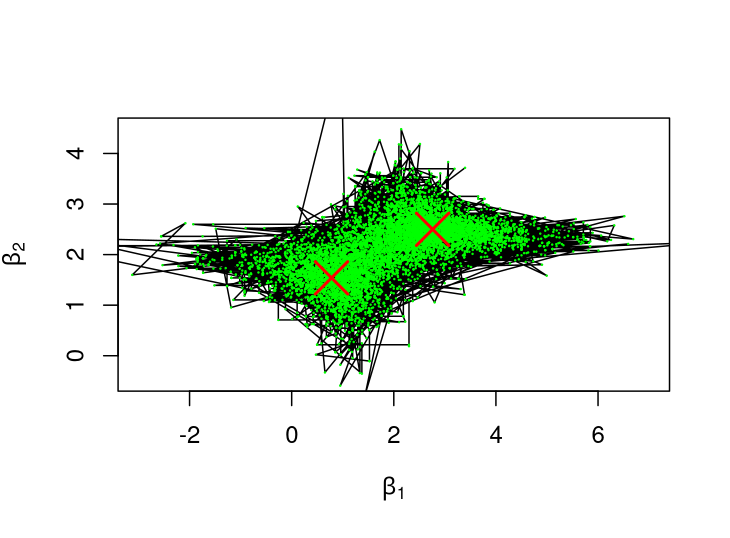} 
    \includegraphics[height=2.5in]{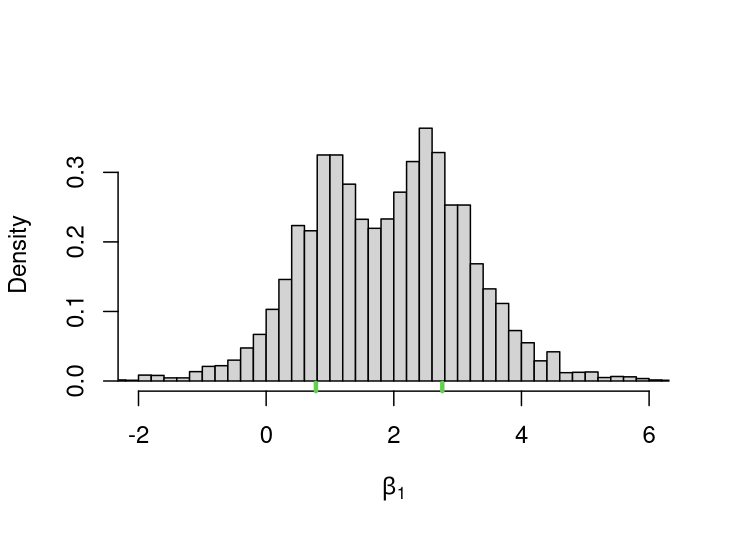} 
    \includegraphics[height=2.5in]{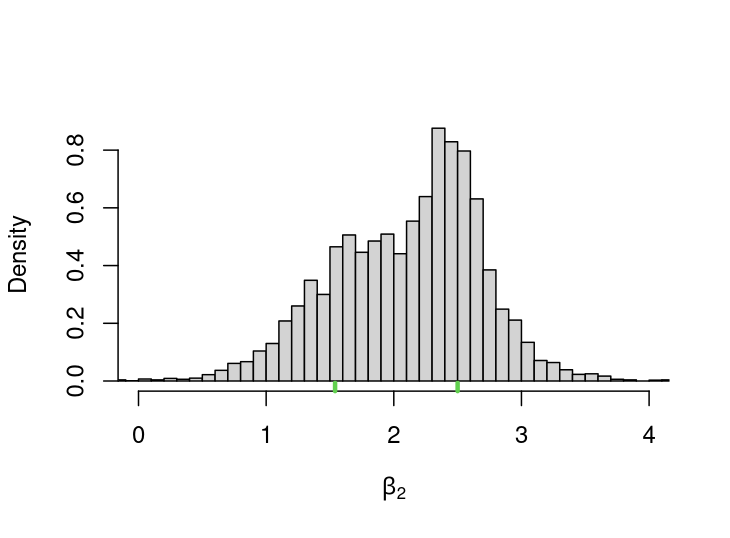} 
\end{center}
\caption{Samples from the adaptive parallel tempering (APT) algorithm for the bivariate seemingly-unrelated regression (SUR) model.}
\label{fig:sur2d}
\end{figure}

\citet{Drton2004} gave an example of a dataset that has two local maxima: $\widehat{\boldsymbol{\beta}}_{1} = (0.78, 1.54)^{\top}$ with $\ell(\widehat{\boldsymbol{\beta}}_{1}) = -27.73$, and $\widehat{\boldsymbol{\beta}}_{2} = (2.76, 2.50)^{\top}$ with $\ell(\widehat{\boldsymbol{\beta}}_{2}) = -27.49$. There is also a saddle point at $(1.62, 2.03)^{\top}$. The values of the regression coefficients differ substantially between the two modes, even though the respective values of the profile log-likelihood are very similar. First, we use \cite{Zellner1962} algorithm to fit the SUR model. This algorithm is implemented in the R package \texttt{systemfit} \citep{Henningsen2007}. It takes 38 iterations to converge to $\widehat{\boldsymbol{\beta}}_{1}$, which is not the global maximum. Furthermore, both of the modes represent a substantial proportion of the probability mass. It is an oversimplification and potentially misleading to only consider one of the modes when estimating $\boldsymbol{\beta}$, since this does not accurately reflect the uncertainty in the parameters. 

We also used the R package \texttt{nimbleAPT} to run the APT algorithm on the same data. It took only 1.09 seconds for 15,000 iterations at 5 temperatures. The first 5,000 iterations were discarded as burn-in. A scatterplot of the remaining 10,000 samples at the target temperature is shown in Figure~\ref{fig:sur2d}, along with one-dimensional marginal histograms.

\section*{Acknowledgements}

K{\L} has been supported by the Royal Society through the Royal Society University Research Fellowship and the Yusuf Hamied Visiting Fellowship. TSF has received funding from the European Research Council (ERC, PrSc-HDBayLe, Grant agreement No. 101076564).

\bibliographystyle{apalike} 
\bibliography{chapter_arxiv}

\begin{thebibliography}{}

\bibitem[Ahn et~al., 2013]{ahn2013distributed}
Ahn, S., Chen, Y., and Welling, M. (2013).
\newblock {Distributed and adaptive darting Monte Carlo through regenerations}.
\newblock In {\em Artificial Intelligence and Statistics}, pages 108--116.

\bibitem[Altekar et~al., 2004]{altekar2004parallel}
Altekar, G., Dwarkadas, S., Huelsenbeck, J.~P., and Ronquist, F. (2004).
\newblock Parallel metropolis coupled markov chain monte carlo for bayesian
  phylogenetic inference.
\newblock {\em Bioinformatics}, 20(3):407--415.

\bibitem[Andricioaei et~al., 2001]{andricioaei2001smart}
Andricioaei, I., Straub, J.~E., and Voter, A.~F. (2001).
\newblock Smart darting {M}onte {C}arlo.
\newblock {\em The Journal of Chemical Physics}, 114(16):6994--7000.

\bibitem[Andrieu and Roberts, 2009]{MR2502648}
Andrieu, C. and Roberts, G.~O. (2009).
\newblock The pseudo-marginal approach for efficient {M}onte {C}arlo
  computations.
\newblock {\em Ann. Statist.}, 37(2):697--725.

\bibitem[Andrieu and Thoms, 2008]{MR2461882}
Andrieu, C. and Thoms, J. (2008).
\newblock A tutorial on adaptive {MCMC}.
\newblock {\em Stat. Comput.}, 18(4):343--373.

\bibitem[Asenjo et~al., 2013]{asenjo2013visualizing}
Asenjo, D., Stevenson, J.~D., Wales, D.~J., and Frenkel, D. (2013).
\newblock Visualizing basins of attraction for different minimization
  algorithms.
\newblock {\em The Journal of Physical Chemistry B}, 117(42):12717--12723.

\bibitem[Atchad{\'e} and Liu, 2010]{atchade2010wang}
Atchad{\'e}, Y.~F. and Liu, J.~S. (2010).
\newblock The {W}ang-{L}andau algorithm in general state spaces: applications
  and convergence analysis.
\newblock {\em Statistica Sinica}, pages 209--233.

\bibitem[Atchad\'{e} and Perron, 2007]{MR2303970}
Atchad\'{e}, Y.~F. and Perron, F. (2007).
\newblock On the geometric ergodicity of {M}etropolis-{H}astings algorithms.
\newblock {\em Statistics}, 41(1):77--84.

\bibitem[Atchad\'e et~al., 2011]{MR2826692}
Atchad\'e, Y.~F., Roberts, G.~O., and Rosenthal, J.~S. (2011).
\newblock Towards optimal scaling of {M}etropolis-coupled {M}arkov chain
  {M}onte {C}arlo.
\newblock {\em Stat. Comput.}, 21(4):555--568.

\bibitem[Bai et~al., 2011]{MR2816538}
Bai, Y., Craiu, R.~V., and Di~Narzo, A.~F. (2011).
\newblock Divide and conquer: a mixture-based approach to regional adaptation
  for {MCMC}.
\newblock {\em J. Comput. Graph. Statist.}, 20(1):63--79.
\newblock Supplementary material available online.

\bibitem[Ballnus et~al., 2017]{ballnus2017comprehensive}
Ballnus, B., Hug, S., Hatz, K., G{\"o}rlitz, L., Hasenauer, J., and Theis,
  F.~J. (2017).
\newblock Comprehensive benchmarking of {M}arkov chain {M}onte {C}arlo methods
  for dynamical systems.
\newblock {\em BMC Systems Biology}, 11:1--18.

\bibitem[Banfield and Raftery, 1992]{banfield1992ice}
Banfield, J.~D. and Raftery, A.~E. (1992).
\newblock Ice floe identification in satellite images using mathematical
  morphology and clustering about principal curves.
\newblock {\em Journal of the American Statistical Association}, 87:7--16.

\bibitem[Baxendale, 2005]{MR2114987}
Baxendale, P.~H. (2005).
\newblock Renewal theory and computable convergence rates for geometrically
  ergodic {M}arkov chains.
\newblock {\em Ann. Appl. Probab.}, 15(1B):700--738.

\bibitem[Bertsimas and Tsitsiklis, 1993]{bertsimas1993simulated}
Bertsimas, D. and Tsitsiklis, J. (1993).
\newblock Simulated annealing.
\newblock {\em Statistical Science}, 8(1):10--15.

\bibitem[Besag, 1974]{besag1974spatial}
Besag, J. (1974).
\newblock Spatial interaction and the statistical analysis of lattice systems.
\newblock {\em Journal of the Royal Statistical Society: Series B
  (Methodological)}, 36(2):192--225.

\bibitem[Beskos et~al., 2008]{MR2444507}
Beskos, A., Roberts, G., Stuart, A., and Voss, J. (2008).
\newblock M{CMC} methods for diffusion bridges.
\newblock {\em Stoch. Dyn.}, 8(3):319--350.

\bibitem[Bornn and Jacob, 2012]{PAWL}
Bornn, L. and Jacob, P.~E. (2012).
\newblock {\em {PAWL}: Implementation of the {PAWL} algorithm}.
\newblock R package version 0.5.

\bibitem[Bornn et~al., 2013]{bornn2013adaptive}
Bornn, L., Jacob, P.~E., Del~Moral, P., and Doucet, A. (2013).
\newblock An adaptive interacting {Wang--Landau} algorithm for automatic
  density exploration.
\newblock {\em Journal of Computational and Graphical Statistics},
  22(3):749--773.

\bibitem[Bouckaert et~al., 2019]{bouckaert2019beast}
Bouckaert, R., Vaughan, T.~G., Barido-Sottani, J., Duch{\^e}ne, S., Fourment,
  M., Gavryushkina, A., Heled, J., Jones, G., K{\"u}hnert, D., De~Maio, N.,
  et~al. (2019).
\newblock {BEAST} 2.5: An advanced software platform for {B}ayesian
  evolutionary analysis.
\newblock {\em PLoS Computational Biology}, 15(4):e1006650.

\bibitem[Chatterjee, 2023]{chatterjee2023spectral}
Chatterjee, S. (2023).
\newblock Spectral gap of nonreversible {M}arkov chains.
\newblock {\em arXiv preprint arXiv:2310.10876}.

\bibitem[Chatterjee and Diaconis, 2013]{chatterjee2013estimating}
Chatterjee, S. and Diaconis, P. (2013).
\newblock Estimating and understanding exponential random graph models.
\newblock {\em Annals of Statistics}, 41(5):2428--2461.

\bibitem[Chimisov et~al., 2018a]{chimisov2018adapting}
Chimisov, C., {\L}atuszy{\'n}ski, K., and Roberts, G. (2018a).
\newblock Adapting the {G}ibbs sampler.
\newblock {\em arXiv preprint arXiv:1801.09299}.

\bibitem[Chimisov et~al., 2018b]{chimisov2018air}
Chimisov, C., Latuszynski, K., and Roberts, G. (2018b).
\newblock Air markov chain monte carlo.
\newblock {\em arXiv e-prints}, pages arXiv--1801.

\bibitem[Chopin and Jacob, 2010]{Chopin2010smc}
Chopin, N. and Jacob, P. (2010).
\newblock Free energy sequential {M}onte {C}arlo, application to mixture
  modelling.
\newblock In {\em Bayesian Statistics}, volume~9, pages 91--118. Oxford
  University Press.

\bibitem[Chopin et~al., 2012]{chopin2012free}
Chopin, N., Leli{\`e}vre, T., and Stoltz, G. (2012).
\newblock Free energy methods for {B}ayesian inference: efficient exploration
  of univariate {G}aussian mixture posteriors.
\newblock {\em Statistics and Computing}, 22:897--916.

\bibitem[Cohn and Fielding, 1999]{MR1700696}
Cohn, H. and Fielding, M. (1999).
\newblock Simulated annealing: searching for an optimal temperature schedule.
\newblock {\em SIAM J. Optim.}, 9(3):779--802.

\bibitem[Cotter et~al., 2013]{MR3135540}
Cotter, S.~L., Roberts, G.~O., Stuart, A.~M., and White, D. (2013).
\newblock M{CMC} methods for functions: modifying old algorithms to make them
  faster.
\newblock {\em Statist. Sci.}, 28(3):424--446.

\bibitem[Craiu et~al., 2009]{MR2750572}
Craiu, R.~V., Rosenthal, J., and Yang, C. (2009).
\newblock Learn from thy neighbor: parallel-chain and regional adaptive {MCMC}.
\newblock {\em J. Amer. Statist. Assoc.}, 104(488):1454--1466.

\bibitem[Dawkins, 1996]{dawkins_wheels}
Dawkins, R. (1996).
\newblock Why don't animals have wheels?
\newblock {\em The Sunday Times}.

\bibitem[{de Valpine} et~al., 2017]{nimble2017}
{de Valpine}, P., Turek, D., Paciorek, C., Anderson-Bergman, C., {Temple Lang},
  D., and Bodik, R. (2017).
\newblock Programming with models: writing statistical algorithms for general
  model structures with {NIMBLE}.
\newblock {\em Journal of Computational and Graphical Statistics}, 26:403--417.

\bibitem[Douc et~al., 2018]{MR3889011}
Douc, R., Moulines, E., Priouret, P., and Soulier, P. (2018).
\newblock {\em Markov chains}.
\newblock Springer Series in Operations Research and Financial Engineering.
  Springer, Cham.

\bibitem[Doucet et~al., 2015]{MR3371005}
Doucet, A., Pitt, M.~K., Deligiannidis, G., and Kohn, R. (2015).
\newblock Efficient implementation of {M}arkov chain {M}onte {C}arlo when using
  an unbiased likelihood estimator.
\newblock {\em Biometrika}, 102(2):295--313.

\bibitem[Drton and Richardson, 2004]{Drton2004}
Drton, M. and Richardson, T.~S. (2004).
\newblock Multimodality of the likelihood in the bivariate seemingly unrelated
  regressions model.
\newblock {\em Biometrika}, 91(2):383--392.

\bibitem[Duane et~al., 1987]{MR3960671}
Duane, S., Kennedy, A.~D., Pendleton, B.~J., and Roweth, D. (1987).
\newblock Hybrid {M}onte {C}arlo.
\newblock {\em Phys. Lett. B}, 195(2):216--222.

\bibitem[Fang et~al., 2014]{fang2014parallel}
Fang, Y., Feng, S., Tam, K.-M., Yun, Z., Moreno, J., Ramanujam, J., and
  Jarrell, M. (2014).
\newblock Parallel tempering simulation of the three-dimensional
  edwards--anderson model with compact asynchronous multispin coding on gpu.
\newblock {\em Computer Physics Communications}, 185(10):2467--2478.

\bibitem[Feroz et~al., 2009]{feroz2009multinest}
Feroz, F., Hobson, M., and Bridges, M. (2009).
\newblock {MultiNest}: an efficient and robust {B}ayesian inference tool for
  cosmology and particle physics.
\newblock {\em Monthly Notices of the Royal Astronomical Society},
  398(4):1601--1614.

\bibitem[Feroz et~al., 2013]{feroz2013importance}
Feroz, F., Hobson, M., Cameron, E., and Pettitt, A. (2013).
\newblock Importance nested sampling and the {MultiNest} algorithm.
\newblock {\em arXiv preprint arXiv:1306.2144}.

\bibitem[Fiebig, 2001]{Fiebig2001}
Fiebig, D.~G. (2001).
\newblock Seemingly unrelated regression.
\newblock In Baltagi, B.~H., editor, {\em A Companion to Theoretical
  Econometrics}. John Wiley \& Sons.

\bibitem[Fort et~al., 2011]{MR3012408}
Fort, G., Moulines, E., and Priouret, P. (2011).
\newblock Convergence of adaptive and interacting {M}arkov chain {M}onte
  {C}arlo algorithms.
\newblock {\em Ann. Statist.}, 39(6):3262--3289.

\bibitem[G{\aa}semyr, 2006]{MR2256484}
G{\aa}semyr, J.~r. (2006).
\newblock The spectrum of the independent {M}etropolis-{H}astings algorithm.
\newblock {\em J. Theoret. Probab.}, 19(1):152--165.

\bibitem[Gelfand and Smith, 1990]{gelfand1990sampling}
Gelfand, A.~E. and Smith, A. F.~M. (1990).
\newblock Sampling-based approaches to calculating marginal densities.
\newblock {\em Journal of the American statistical association},
  85(410):398--409.

\bibitem[Geyer, 1991]{geyer1991markov}
Geyer, C.~J. (1991).
\newblock Markov chain {M}onte {C}arlo maximum likelihood.
\newblock {\em {Computing Science and Statistics}}, 23:156--163.

\bibitem[Gon{\c{c}}alves et~al., 2023]{gonccalves2023exact}
Gon{\c{c}}alves, F.~B., {\L}atuszy{\'n}ski, K., and Roberts, G.~O. (2023).
\newblock Exact {M}onte {C}arlo likelihood-based inference for jump-diffusion
  processes.
\newblock {\em Journal of the Royal Statistical Society Series B: Statistical
  Methodology}, 85(3):732--756.

\bibitem[Greene, 1997]{Greene1997}
Greene, W.~H. (1997).
\newblock {\em Econometric Analysis}.
\newblock Prentice Hall, 3$^\mathrm{rd}$ edition.

\bibitem[Grenon-Godbout and B\'{e}dard, 2021]{MR4267920}
Grenon-Godbout, N. and B\'{e}dard, M. (2021).
\newblock On-line partitioning of the sample space in the regional adaptive
  algorithm.
\newblock {\em Canad. J. Statist.}, 49(2):238--261.

\bibitem[Haario et~al., 2001]{MR1828504}
Haario, H., Saksman, E., and Tamminen, J. (2001).
\newblock An adaptive metropolis algorithm.
\newblock {\em Bernoulli}, 7(2):223--242.

\bibitem[Henningsen and Hamann, 2007]{Henningsen2007}
Henningsen, A. and Hamann, J.~D. (2007).
\newblock {systemfit}: A package for estimating systems of simultaneous
  equations in {R}.
\newblock {\em Journal of Statistical Software}, 23(4):1--40.

\bibitem[Higdon, 1998]{higdon1998auxiliary}
Higdon, D.~M. (1998).
\newblock Auxiliary variable methods for {M}arkov chain {M}onte {C}arlo with
  applications.
\newblock {\em Journal of the American Statistical Association},
  93(442):585--595.

\bibitem[Ihler et~al., 2004]{ihler2004nonparametric}
Ihler, A.~T., Fisher~III, J.~W., Moses, R.~L., and Willsky, A.~S. (2004).
\newblock Nonparametric belief propagation for self-calibration in sensor
  networks.
\newblock In {\em Proceedings of the 3rd international symposium on Information
  processing in sensor networks}, pages 225--233.

\bibitem[Ising, 1925]{ising1925beitrag}
Ising, E. (1925).
\newblock Beitrag zur theorie des ferromagnetismus.
\newblock {\em Zeit. fur Physik}, 31:253--258.

\bibitem[Jasra et~al., 2005]{MR2182987}
Jasra, A., Holmes, C.~C., and Stephens, D.~A. (2005).
\newblock Markov chain {M}onte {C}arlo methods and the label switching problem
  in {B}ayesian mixture modeling.
\newblock {\em Statist. Sci.}, 20(1):50--67.

\bibitem[Jerrum and Sinclair, 1988]{jerrum1988conductance}
Jerrum, M. and Sinclair, A. (1988).
\newblock Conductance and the rapid mixing property for {M}arkov chains: the
  approximation of permanent resolved.
\newblock In {\em Proceedings of the twentieth annual ACM symposium on Theory
  of computing}, pages 235--244.

\bibitem[Kamberaj, 2020]{kamberaj2020molecular}
Kamberaj, H. (2020).
\newblock {\em Molecular dynamics simulations in statistical physics: theory
  and applications}.
\newblock Springer.

\bibitem[Kingman, 1982]{MR0671034}
Kingman, J. F.~C. (1982).
\newblock The coalescent.
\newblock {\em Stochastic Process. Appl.}, 13(3):235--248.

\bibitem[Kirkpatrick et~al., 1983]{MR0702485}
Kirkpatrick, S., Gelatt, Jr., C.~D., and Vecchi, M.~P. (1983).
\newblock Optimization by simulated annealing.
\newblock {\em Science}, 220(4598):671--680.

\bibitem[Kontoyiannis and Meyn, 2012]{MR2981426}
Kontoyiannis, I. and Meyn, S.~P. (2012).
\newblock Geometric ergodicity and the spectral gap of non-reversible {M}arkov
  chains.
\newblock {\em Probab. Theory Related Fields}, 154(1-2):327--339.

\bibitem[{\L}{\k{a}}cki and Miasojedow, 2016]{lkacki2016state}
{\L}{\k{a}}cki, M.~K. and Miasojedow, B. (2016).
\newblock State-dependent swap strategies and automatic reduction of number of
  temperatures in adaptive parallel tempering algorithm.
\newblock {\em Statistics and Computing}, 26:951--964.

\bibitem[Lan et~al., 2014]{lan2014wormhole}
Lan, S., Streets, J., and Shahbaba, B. (2014).
\newblock Wormhole {H}amiltonian {M}onte {C}arlo.
\newblock In {\em Proceedings of the AAAI Conference on Artificial
  Intelligence}, volume~28.

\bibitem[Lawler and Sokal, 1988]{MR0930082}
Lawler, G.~F. and Sokal, A.~D. (1988).
\newblock Bounds on the {$L^2$} spectrum for {M}arkov chains and {M}arkov
  processes: a generalization of {C}heeger's inequality.
\newblock {\em Trans. Amer. Math. Soc.}, 309(2):557--580.

\bibitem[Levin and Peres, 2017]{MR3726904}
Levin, D.~A. and Peres, Y. (2017).
\newblock {\em Markov chains and mixing times}.
\newblock American Mathematical Society, Providence, RI, second edition.
\newblock With contributions by Elizabeth L. Wilmer, With a chapter on
  ``Coupling from the past'' by James G. Propp and David B. Wilson.

\bibitem[Liang, 2005]{liang2005generalized}
Liang, F. (2005).
\newblock A generalized {W}ang--{L}andau algorithm for {M}onte {C}arlo
  computation.
\newblock {\em Journal of the American Statistical Association},
  100(472):1311--1327.

\bibitem[Lingenheil et~al., 2009]{lingenheil2009efficiency}
Lingenheil, M., Denschlag, R., Mathias, G., and Tavan, P. (2009).
\newblock Efficiency of exchange schemes in replica exchange.
\newblock {\em Chemical Physics Letters}, 478(1-3):80--84.

\bibitem[Liu, 1996]{liu1996metropolized}
Liu, J.~S. (1996).
\newblock Metropolized independent sampling with comparisons to rejection
  sampling and importance sampling.
\newblock {\em Statistics and Computing}, 6(2):113--119.

\bibitem[Madras and Randall, 2002]{MR1910641}
Madras, N. and Randall, D. (2002).
\newblock Markov chain decomposition for convergence rate analysis.
\newblock {\em Ann. Appl. Probab.}, 12(2):581--606.

\bibitem[Madras and Zheng, 2003]{MR1943860}
Madras, N. and Zheng, Z. (2003).
\newblock On the swapping algorithm.
\newblock {\em Random Structures Algorithms}, 22(1):66--97.

\bibitem[Marinari and Parisi, 1992]{marinari1992simulated}
Marinari, E. and Parisi, G. (1992).
\newblock Simulated tempering: a new {M}onte {C}arlo scheme.
\newblock {\em Europhysics Letters}, 19(6):451.

\bibitem[Meyn and Tweedie, 1993]{MR1287609}
Meyn, S.~P. and Tweedie, R.~L. (1993).
\newblock {\em Markov chains and stochastic stability}.
\newblock Communications and Control Engineering Series. Springer-Verlag
  London, Ltd., London.

\bibitem[Miasojedow et~al., 2013]{Miasojedow2013}
Miasojedow, B., Moulines, E., and Vihola, M. (2013).
\newblock An adaptive parallel tempering algorithm.
\newblock {\em Journal of Computational and Graphical Statistics},
  22(3):649--664.

\bibitem[Mihail, 1989]{mihail1989conductance}
Mihail, M. (1989).
\newblock Conductance and convergence of {M}arkov chains --- a combinatorial
  treatment of expanders.
\newblock In {\em 30th Annual Symposium on Foundations of Computer Science},
  pages 526--531. IEEE Computer Society.

\bibitem[Mingas and Bouganis, 2012]{mingas2012parallel}
Mingas, G. and Bouganis, C.-S. (2012).
\newblock Parallel tempering mcmc acceleration using reconfigurable hardware.
\newblock In {\em International Symposium on Applied Reconfigurable Computing},
  pages 227--238. Springer.

\bibitem[Mira, 2001]{MR1888449}
Mira, A. (2001).
\newblock Ordering and improving the performance of {M}onte {C}arlo {M}arkov
  chains.
\newblock {\em Statist. Sci.}, 16(4):340--350.

\bibitem[Neal, 1999]{MR1723510}
Neal, R.~M. (1999).
\newblock Regression and classification using {G}aussian process priors.
\newblock In {\em Bayesian statistics}, volume~6, pages 475--501. Oxford
  University Press.

\bibitem[Neal, 2011]{neal2011mcmc}
Neal, R.~M. (2011).
\newblock {MCMC} using {H}amiltonian dynamics.
\newblock In {\em Handbook of Markov Chain Monte Carlo}, chapter~5, pages
  113--162. Chapman \& Hall/CRC Press.

\bibitem[Niemiro, 1995]{MR1377584}
Niemiro, W. (1995).
\newblock Limit distributions of simulated annealing {M}arkov chains.
\newblock {\em Discuss. Math. Algebra Stochastic Methods}, 15(2):241--269.

\bibitem[Okabe et~al., 2001]{okabe2001replica}
Okabe, T., Kawata, M., Okamoto, Y., and Mikami, M. (2001).
\newblock Replica-exchange monte carlo method for the isobaric--isothermal
  ensemble.
\newblock {\em Chemical physics letters}, 335(5-6):435--439.

\bibitem[Papamarkou et~al., 2022]{MR4444376}
Papamarkou, T., Hinkle, J., Young, M.~T., and Womble, D. (2022).
\newblock Challenges in {M}arkov chain {M}onte {C}arlo for {B}ayesian neural
  networks.
\newblock {\em Statist. Sci.}, 37(3):425--442.

\bibitem[Pleydell, 2021]{nimbleAPT}
Pleydell, D.~R. (2021).
\newblock {\em {nimbleAPT}: Adaptive Parallel Tempering with NIMBLE.}
\newblock {R} package version 1.0.4.

\bibitem[Pompe et~al., 2020]{Pompe2018}
Pompe, E., Holmes, C., and {\L}atuszy{\'n}ski, K. (2020).
\newblock A framework for adaptive {MCMC} targeting multimodal distributions.
\newblock {\em The Annals of Statistics}, 48(5):2930--2952.

\bibitem[Roberts et~al., 1997]{MR1428751}
Roberts, G.~O., Gelman, A., and Gilks, W.~R. (1997).
\newblock Weak convergence and optimal scaling of random walk {M}etropolis
  algorithms.
\newblock {\em Ann. Appl. Probab.}, 7(1):110--120.

\bibitem[Roberts and Rosenthal, 1997]{MR1448322}
Roberts, G.~O. and Rosenthal, J.~S. (1997).
\newblock Geometric ergodicity and hybrid {M}arkov chains.
\newblock {\em Electron. Comm. Probab.}, 2:no. 2, 13--25.

\bibitem[Roberts and Rosenthal, 1998a]{MR1624414}
Roberts, G.~O. and Rosenthal, J.~S. (1998a).
\newblock Markov-chain {M}onte {C}arlo: some practical implications of
  theoretical results.
\newblock {\em Canad. J. Statist.}, 26(1):5--31.
\newblock With discussion by Hemant Ishwaran and Neal Madras and a rejoinder by
  the authors.

\bibitem[Roberts and Rosenthal, 1998b]{MR1625691}
Roberts, G.~O. and Rosenthal, J.~S. (1998b).
\newblock Optimal scaling of discrete approximations to {L}angevin diffusions.
\newblock {\em J. R. Stat. Soc. Ser. B Stat. Methodol.}, 60(1):255--268.

\bibitem[Roberts and Rosenthal, 2001]{MR1888450}
Roberts, G.~O. and Rosenthal, J.~S. (2001).
\newblock Optimal scaling for various {M}etropolis-{H}astings algorithms.
\newblock {\em Statist. Sci.}, 16(4):351--367.

\bibitem[Roberts and Rosenthal, 2004]{MR2095565}
Roberts, G.~O. and Rosenthal, J.~S. (2004).
\newblock General state space {M}arkov chains and {MCMC} algorithms.
\newblock {\em Probab. Surv.}, 1:20--71.

\bibitem[Roberts and Rosenthal, 2007]{MR2340211}
Roberts, G.~O. and Rosenthal, J.~S. (2007).
\newblock Coupling and ergodicity of adaptive {M}arkov chain {M}onte {C}arlo
  algorithms.
\newblock {\em J. Appl. Probab.}, 44(2):458--475.

\bibitem[Roberts and Rosenthal, 2009]{MR2749836}
Roberts, G.~O. and Rosenthal, J.~S. (2009).
\newblock Examples of adaptive {MCMC}.
\newblock {\em J. Comput. Graph. Statist.}, 18(2):349--367.

\bibitem[Roberts et~al., 2022]{Roberts2020}
Roberts, G.~O., Rosenthal, J.~S., and Tawn, N.~G. (2022).
\newblock Skew {B}rownian motion and complexity of the {ALPS} algorithm.
\newblock {\em Journal of Applied Probability}, 59(3):777--796.

\bibitem[Roberts and Sahu, 1997]{roberts1997updating}
Roberts, G.~O. and Sahu, S.~K. (1997).
\newblock Updating schemes, correlation structure, blocking and
  parameterization for the {G}ibbs sampler.
\newblock {\em Journal of the Royal Statistical Society Series B: Statistical
  Methodology}, 59(2):291--317.

\bibitem[Roberts and Tweedie, 1996]{MR1440273}
Roberts, G.~O. and Tweedie, R.~L. (1996).
\newblock Exponential convergence of {L}angevin distributions and their
  discrete approximations.
\newblock {\em Bernoulli}, 2(4):341--363.

\bibitem[Rosenthal, 2011]{rosenthal2011optimal}
Rosenthal, J.~S. (2011).
\newblock Optimal proposal distributions and adaptive {MCMC}.
\newblock In {\em Handbook of Markov Chain Monte Carlo}, chapter~4, pages
  93--112. Chapman \& Hall/CRC Press.

\bibitem[Searle and Khuri, 2017]{Searle2017}
Searle, S.~R. and Khuri, A.~I. (2017).
\newblock {\em Matrix Algebra Useful for Statistics}.
\newblock John Wiley \& Sons, 2$^\mathrm{nd}$ edition.

\bibitem[Shaby and Wells, 2011]{shaby2010exploring}
Shaby, B. and Wells, M.~T. (2011).
\newblock Exploring an adaptive {M}etropolis algorithm.
\newblock {\em Department of Statistics, Duke University}, 2011-14.

\bibitem[Sherlock et~al., 2015]{MR3285606}
Sherlock, C., Thiery, A.~H., Roberts, G.~O., and Rosenthal, J.~S. (2015).
\newblock On the efficiency of pseudo-marginal random walk {M}etropolis
  algorithms.
\newblock {\em Ann. Statist.}, 43(1):238--275.

\bibitem[Sminchisescu and Welling, 2011]{sminchisescu2011generalized}
Sminchisescu, C. and Welling, M. (2011).
\newblock {Generalized darting Monte Carlo}.
\newblock {\em Pattern Recognition}, 44(10):2738--2748.

\bibitem[Smith and Roberts, 1993]{smith1993bayesian}
Smith, A. F.~M. and Roberts, G.~O. (1993).
\newblock Bayesian computation via the {G}ibbs sampler and related {M}arkov
  chain {M}onte {C}arlo methods.
\newblock {\em Journal of the Royal Statistical Society: Series B
  (Methodological)}, 55(1):3--23.

\bibitem[Srivastava and Giles, 1987]{Srivastava1987}
Srivastava, V.~K. and Giles, D. E.~A. (1987).
\newblock {\em Seemingly Unrelated Regression Equation Models: Estimation and
  Inference}.
\newblock Marcel Dekker.

\bibitem[Sundberg, 2010]{Sundberg2010}
Sundberg, R. (2010).
\newblock Flat and multimodal likelihoods and model lack of fit in curved
  exponential families.
\newblock {\em Scandinavian Journal of Statistics}, 37(4):632--643.

\bibitem[Swendsen and Wang, 1986]{MR0869788}
Swendsen, R.~H. and Wang, J.-S. (1986).
\newblock Replica {M}onte {C}arlo simulation of spin-glasses.
\newblock {\em Phys. Rev. Lett.}, 57(21):2607--2609.

\bibitem[Syed et~al., 2022]{syed2022non}
Syed, S., Bouchard-C{\^o}t{\'e}, A., Deligiannidis, G., and Doucet, A. (2022).
\newblock Non-reversible parallel tempering: a scalable highly parallel {MCMC}
  scheme.
\newblock {\em Journal of the Royal Statistical Society Series B: Statistical
  Methodology}, 84(2):321--350.

\bibitem[Tak et~al., 2018]{Tak2016}
Tak, H., Meng, X.-L., and van Dyk, D.~A. (2018).
\newblock A repelling--attracting {M}etropolis algorithm for multimodality.
\newblock {\em Journal of Computational and Graphical Statistics},
  27(3):479--490.

\bibitem[Tawn et~al., 2021]{tawn2021alps}
Tawn, N.~G., Moores, M.~T., and Roberts, G.~O. (2021).
\newblock Annealed leap-point sampler for multimodal target distributions.
\newblock {\em arXiv preprint arXiv:2112.12908}.

\bibitem[Tawn and Roberts, 2019]{Tawn2018}
Tawn, N.~G. and Roberts, G.~O. (2019).
\newblock Accelerating parallel tempering: Quantile tempering algorithm
  {(QuanTA)}.
\newblock {\em Advances in Applied Probability}, 51(3):802--834.

\bibitem[Tawn et~al., 2020]{GarethJeffNick}
Tawn, N.~G., Roberts, G.~O., and Rosenthal, J.~S. (2020).
\newblock Weight-preserving simulated tempering.
\newblock {\em Statistics and Computing}, 30(1):27--41.

\bibitem[Tjelmeland and Eidsvik, 2004]{Tjelmeland2004}
Tjelmeland, H. and Eidsvik, J. (2004).
\newblock On the use of local optimizations within {M}etropolis–{H}astings
  updates.
\newblock {\em Journal of the Royal Statistical Society: Series B (Statistical
  Methodology)}, 66(2):411--427.

\bibitem[Tjelmeland and Hegstad, 2001]{MR1844357}
Tjelmeland, H. and Hegstad, B.~K. (2001).
\newblock Mode jumping proposals in {MCMC}.
\newblock {\em {Scandinavian Journal of Statistics}}, 28(1):205--223.

\bibitem[van~der Vaart, 1998]{MR1652247}
van~der Vaart, A.~W. (1998).
\newblock {\em Asymptotic statistics}, volume~3 of {\em Cambridge Series in
  Statistical and Probabilistic Mathematics}.
\newblock Cambridge University Press, Cambridge.

\bibitem[Vats et~al., 2022]{MR4430963}
Vats, D., Gon\c{c}alves, F.~B., {\L}atuszy\'{n}ski, K., and Roberts, G.~O.
  (2022).
\newblock Efficient {B}ernoulli factory {M}arkov chain {M}onte {C}arlo for
  intractable posteriors.
\newblock {\em Biometrika}, 109(2):369--385.

\bibitem[Vishwanath and Tak, 2024]{Rep_Att_HMC}
Vishwanath, S. and Tak, H. (2024).
\newblock Repelling-attracting {H}amiltonian {M}onte {C}arlo.
\newblock {\em arXiv preprint arXiv:2403.04607}.

\bibitem[Vu et~al., 2023]{vu2023warped}
Vu, Q., Moores, M.~T., and {Zammit-Mangion}, A. (2023).
\newblock Warped gradient-enhanced {G}aussian process surrogate models for
  exponential family likelihoods with intractable normalizing constants.
\newblock {\em Bayesian Analysis}, to appear.

\bibitem[Wang and Landau, 2001]{Wang2001a}
Wang, F. and Landau, D.~P. (2001).
\newblock Determining the density of states for classical statistical models: A
  random walk algorithm to produce a flat histogram.
\newblock {\em {Physical Review E}}, 64(5):056101.

\bibitem[Wang et~al., 2023]{MR4675034}
Wang, Y., Polson, N., and Sokolov, V.~O. (2023).
\newblock Data augmentation for {B}ayesian deep learning.
\newblock {\em Bayesian Anal.}, 18(4):1041--1069.

\bibitem[Woodard et~al., 2009a]{woodard2009conditions}
Woodard, D.~B., Schmidler, S.~C., and Huber, M. (2009a).
\newblock Conditions for rapid mixing of parallel and simulated tempering on
  multimodal distributions.
\newblock {\em {The Annals of Applied Probability}}, pages 617--640.

\bibitem[Woodard et~al., 2009b]{woodard2009sufficient}
Woodard, D.~B., Schmidler, S.~C., and Huber, M. (2009b).
\newblock Sufficient conditions for torpid mixing of parallel and simulated
  tempering.
\newblock {\em {Electronic Journal of Probability}}, 14:780--804.

\bibitem[Zellner, 1962]{Zellner1962}
Zellner, A. (1962).
\newblock An efficient method of estimating seemingly unrelated regressions and
  tests for aggregation bias.
\newblock {\em Journal of the American Statistical Association},
  57(298):348--368.

\bibitem[Zhou, 2011]{zhou2011multi}
Zhou, Q. (2011).
\newblock Multi-domain sampling with applications to structural inference of
  {B}ayesian networks.
\newblock {\em Journal of the American Statistical Association},
  106(496):1317--1330.

\end{thebibliography}

\end{document}